\documentclass[12pt,preprint,letterpaper]{aastex}
\pdfoutput=1
\slugcomment{Accepted for publication in the Astronomical Journal -- 2009 Mar 19}

\newcommand\sst{\textit{Spitzer Space Telescope}}
\newcommand\spitzer{\textit{Spitzer}}
\newcommand\hst{\textit{Hubble Space Telescope}}
\newcommand\rosetta{\textit{Rosetta}}

\newcommand\inv{$^{-1}$}
\newcommand\mjysr{MJy~sr$^{-1}$}
\newcommand\invrho{~$\rho^{-1}$}
\newcommand\kms{km~s$^{-1}$}

\newcommand\gcm{g~cm$^{-3}$}

\newcommand\blnk{\multicolumn{2}{c}{\nodata}}

\begin{document}

\title{\spitzer{} Observations of Comet 67P/Churyumov-Gerasimenko at
  5.5--4.3~AU From the Sun}

\shorttitle{Comet 67P at 5.5--4.3~AU}

\author{Michael S. Kelley,\altaffilmark{1,2} Diane
  H. Wooden,\altaffilmark{3} Cecilia Tubiana\altaffilmark{4}, Hermann
  Boehnhardt,\altaffilmark{4} Charles E. Woodward,\altaffilmark{5}
  David E. Harker,\altaffilmark{6}}

\altaffiltext{1}{Department of Physics, University of Central Florida,
  4000 Central Florida Blvd., Orlando, FL, 32816-2385}

\altaffiltext{2}{Current address: Department of Astronomy, University
  of Maryland, College Park, MD, 20742-2421; msk@astro.umd.edu}

\altaffiltext{3}{NASA Ames Research Center, Space Science Division, MS
  245-3, Moffett Field, CA 94035; Diane.H.Wooden@nasa.gov}

\altaffiltext{4}{Max Planck Institute for Solar System Research,
  Max-Planck-Str. 2, 37191 Katlenburg-Lindau, Germany;
  boehnhardt@mps.mpg.de, tubiana@mps.mpg.de}

\altaffiltext{5}{Department of Astronomy, University of Minnesota, 116
  Church St SE, Minneapolis, MN, 55455; chelsea@astro.umn.edu}

\altaffiltext{6}{Center for Astronomy and Space Science, University of
  California-San Diego, 9500 Gilman Drive, Department 0424, San Diego,
  CA 92093; dharker@ucsd.edu}

\begin{abstract}
We report \sst{} observations of comet 67P/Churyumov-Gerasimenko at
5.5 and 4.3~AU from the Sun, post-aphelion.  Comet 67P is the primary
target of the European Space Agency's \textit{Rosetta} mission.  The
\rosetta{} spacecraft will rendezvous with the nucleus at heliocentric
distances similar to our observations.  Rotationally resolved
observations at 8 and 24~\micron{} (at a heliocentric distance, $r_h$,
of 4.8~AU) that sample the size and color-temperature of the nucleus
are combined with aphelion $R$-band light curves observed at the Very
Large Telescope (VLT) and yield a mean effective radius of
$2.04\pm0.11$~km, and an $R$-band geometric albedo of $0.054\pm0.006$.
The amplitudes of the $R$-band and mid-infrared light curves agree,
which suggests that the variability is dominated by the shape of the
nucleus.  We also detect the dust trail of the comet at 4.8 and
5.5~AU, constrain the grain sizes to be $\lesssim6$~mm, and estimate
the impact hazard to \rosetta.  We find no evidence for recently
ejected dust in our images.  If the activity of 67P is consistent from
orbit to orbit, then we may expect the \rosetta{} spacecraft will
return images of an inactive or weakly active nucleus as it rendezvous
with the comet at $r_h=4$~AU in 2014.

\end{abstract}


\section{INTRODUCTION}

Ecliptic comet 67P/Churyumov-Gerasimenko (67P) is the primary target
of the \rosetta{} mission, which will rendezvous with the nucleus in
2014 when the comet has passed aphelion and is 4~AU from the Sun
(aphelion distance 5.71~AU).  \rosetta{} will orbit with the comet
through perihelion in 2015 (perihelion distance 1.25~AU).  The
\rosetta{} spacecraft is designed to characterize the comet nucleus
(morphology, composition) and coma (development of activity, dust-gas
interaction, interaction with the solar wind), with both orbiting and
landing spacecraft \citep{glassmeier07}.  Understanding the comet's
gas and dust behavior near aphelion is important for mission planning.

In previous perihelion passages, the comet was not characterized over
the orbital arc at which the \rosetta{} spacecraft will rendezvous
with the nucleus \citep{agarwal07b}.  Dynamical models by
\citet{fulle04} of the dust coma and tail of 67P observed near
perihelion in 2002--2003 have indicated that the comet is
significantly active and able to eject millimeter and centimeter sized
grains at 3.6~AU, post-aphelion (i.e., pre-perihelion).  However, this
conclusion is disputed by \citet{agarwal07b}, who point out that the
dust tail analyzed by \citet{fulle04} contains overlaping
contributions from all dust ejected before the comet reached 1.5~AU.
Direct imaging of the comet with ground-based observations have shown
that the comet is active at 2.0~AU, post-aphelion \citep{agarwal07b}.
Recent ground-based optical and \sst{} mid-infrared (mid-IR)
observations have extended that limit to 2.9--3.0~AU \citep{kadota08,
  wooden08}.  Some ecliptic comets are active at more than 4~AU from
the Sun, e.g., comets 65P/Gunn, 74P/Smirnova-Chernykh,
152P/Helin-Lawrence, and 103P/Hartley \citep{lowry01, lowry99,
  snodgrass08}.  Indeed, even comet 67P was brighter than expected for
a bare nucleus and dust trail in \spitzer{} spectra of the comet at a
heliocentric distance ($r_h$) of 4.98~AU, pre-aphelion
\citep{kelley06-apj}, and in ground-based CCD observations at 4.9~AU,
pre-aphelion \citep{mueller92, lamy06}---equivalent post-aphelion
observations have only recently been presented \citep{tubiana08}.  The
present study of comet 67P is motivated by the lack of aphelion
observations and the importance of the characterization of the nucleus
and comet activity of 67P for the \rosetta{} mission.

In this paper, we present and discuss \sst{} \citep{gehrz07, werner04}
observations of comet 67P at heliocentric distances between 5.5 and
4.3~AU, post-aphelion---the latter distance is comparable to the $r_h$
at which \rosetta{} will rendezvous with the comet nucleus.  In \S2,
we present imaging observations of the comet obtained with both the
Multiband Imaging Photometer for \spitzer{} \citep[MIPS;][]{rieke04}
at 24~\micron{} and the Infrared Array Camera \citep[IRAC;][]{fazio04}
at 8~\micron{}, and 14--35~\micron{} spectra of the nucleus with the
Infrared Spectrograph \citep[IRS;][]{houck04}.  In 24~\micron{} MIPS
images, we detect the comet's dust trail \citep{sykes92} and a tail
composed of large grains.  We constrain the largest grain size ejected
by 67P to be $\lesssim6$~mm, and then estimate the large grain number
density and its impact hazard to \rosetta{} in \S\ref{sec:trail}.  Our
observations are consistent with an inactive nucleus between 5.5 and
4.3~AU, post-aphelion.  We present upper-limits to emission from dust
at 5.5 and 4.8~AU in \S\ref{sec:coma}.  In \S\ref{sec:nucleus}, we
derive the infrared (IR) light curve at 4.8~AU by sampling the
emission from the nucleus at 11 epochs with consecutive MIPS and IRAC
observations.  The \spitzer{} derived light curves of comet 67P
constrain the shape, color-temperature, and effective radius of the
nucleus.  We examine those properties and their behavior with
rotational phase to show in \S\ref{sec:lightcurve} that the visible
light curve is likely dominated by nucleus shape rather than albedo.
However, there are some discrepancies between the visible and IR light
curves, therefore some albedo variation cannot be ruled out.  Our
results are summarized in \S\ref{sec:summary}.

\section{OBSERVATIONS AND REDUCTION}\label{sec:obs}

\subsection{Imaging}
\begin{figure}
\includegraphics[width=\columnwidth]{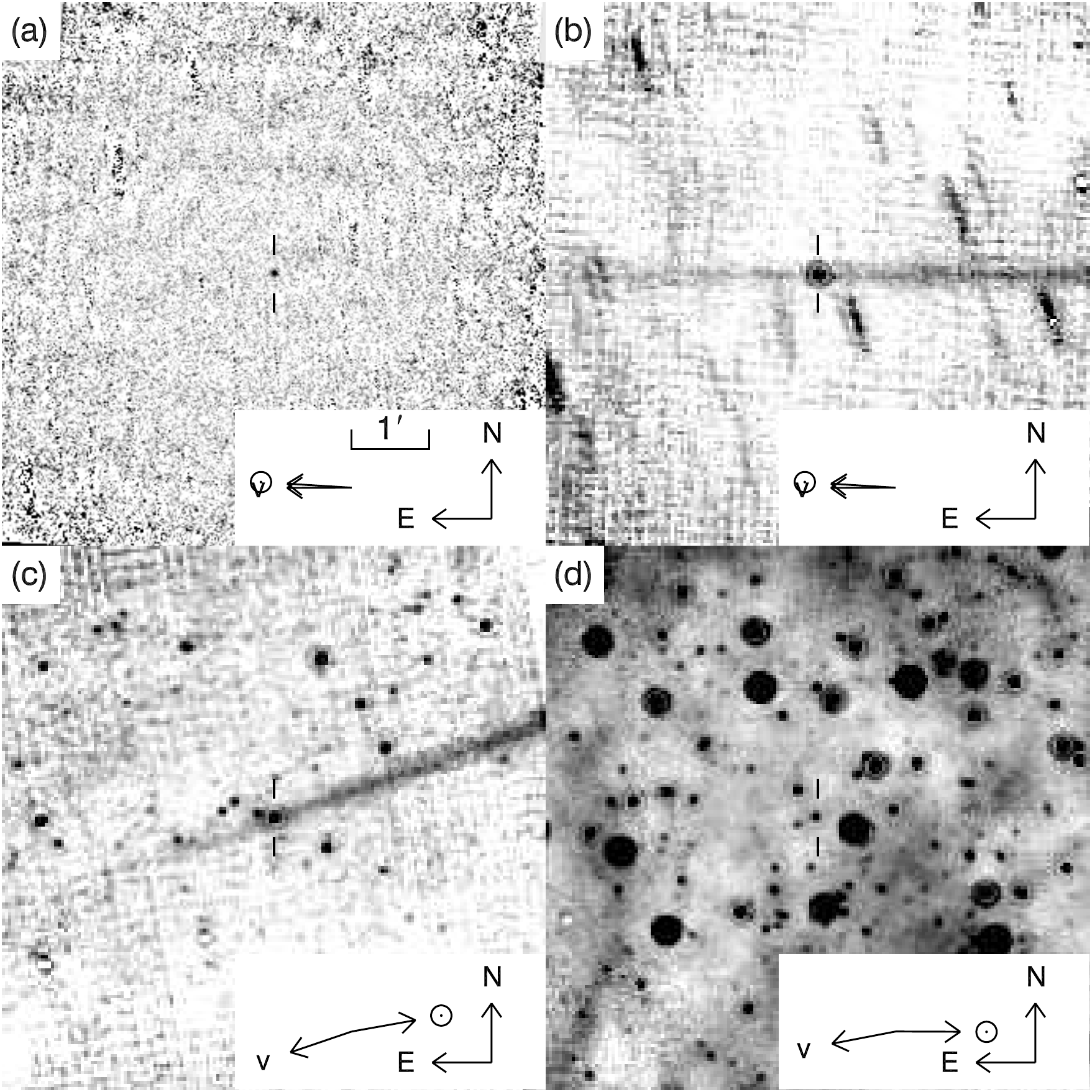}
\caption{\spitzer{} images of comet 67P/Churyumov-Gerasimenko.  The
  nucleus is located at the center of all images and marked with
  \textit{vertical bars} for clarity.  The image scale is provided in
  (a).  North, east, the projected Sun angle ($\sun$), and the
  projected comet velocity vector (v) are indicated in the lower-right
  of each image: (a) a co-added mosaic of all 11 IRAC 8~\micron{}
  observations from 2007 May 17 (4.8~AU); (b) a co-added mosaic of all
  11 MIPS 24~\micron{} images from 2007 May 18 (4.8~AU); (c) the MIPS
  24~\micron{} image taken on 2006 September 01 (5.5~AU); and (d) a
  MIPS 24~\micron{} image from 2007 September 19 (4.4~AU).  The dust
  trail crosses two of the MIPS images (b and c) along the velocity
  vector.  \label{fig:images}}
\end{figure}

\spitzer{} observed comet 67P with the MIPS 24~\micron{} instrument in
Photometry/Super Resolution mode (pixel scale
2.55~arcsec~pixel$^{-1}$).  This mode provides 14 dithered images
covering an approximate field-of-view of $7\arcmin\times8\arcmin$
(each image is approximately $5\arcmin\times5\arcmin$).  The comet
nucleus is centered in the dither pattern, and the telescope tracks
the comet with its computed non-sidereal rates.  Each image has an
exposure time of 10~s.  The dither pattern was executed 9 times for
each of 2 epochs at $r_h=5.5$~AU (2006 August 16 and 2006 September
01), 3 times for each of 11 epochs at 4.8~AU (2007 May 18), and once
for each of 3 epochs at 4.4~AU (2007 September 19).  The comet is
readily identified as a moving point source, centered in the
field-of-view, and bisected by the comet's dust emission
(Fig.~\ref{fig:images}), except at 4.4~AU where dust is not observed
due to crowding by point sources and diffuse emission arising from the
interstellar medium.  For the 5.5 and 4.8~AU epochs, we median
combined each observation's images and divided each image with the
result to remove image artifacts \citep{mdh}---the 4.4~AU images are
too crowed for this step.  We stacked the images together in the rest
frame of the comet with the MOPEX software \citep{makovoz05} to
produce 2 mosaics at 5.5~AU, 11 at 4.8~AU, and 3 at 4.4~AU.  The 11
MIPS 24~\micron{} images at 4.8~AU spanned a duration of 13.17~hr
(start to finish) enabling sampling of the light curve over
approximately one rotational period of the nucleus
($P_{rot}\approx13$~hr, see \S\ref{sec:nucleus}).
Table~\ref{table:data} presents the observation parameters for the
MIPS data.

\noindent{\textbf{[Table \ref{table:data}, Figure \ref{fig:images}]}}

We also observed comet 67P with the IRAC 8~\micron{} camera in mapping
mode (pixel scale 1.22~arcsec~pixel$^{-1}$).  The 11 observations
spanned a duration of 13.09~hr (start to finish) and, similar to the
MIPS observations, measured the light curve just over one rotational
period of the nucleus at $r_h=4.8$~AU (2007 May 17).  The observations
were designed with a 14-point cyclic dither pattern, and 30~s exposure
times.  The images of each observation are mosaicked in the rest frame
of the comet with the MOPEX software.  The nucleus is faint at
8~\micron{} and not readily identifiable.  We median combined all 11
images and found the comet to be near the center of the frame
(Fig.~\ref{fig:images}).  The comet was not detected with the IRAC
4.5~\micron{} camera, as expected for a $R=2$~km nucleus at this
distance.  Table~\ref{table:data} presents the observation parameters
for the IRAC data.

Because the IRAC and MIPS instruments are not simultaneously operated,
\spitzer{} observed the two light curves as close in time as was
operationally feasible.  The IRAC observations were executed first,
one rotational period was skipped for instrument calibration and
changeover, then the MIPS observations were obtained.  We assumed a
rotational period of 12.8--12.9~hr to constrain each light curve using
approximately equal time steps between each observation.  This allows
us to phase the two wavelengths (8 and 24~\micron) together without
interpolation.  \citet{tubiana08} recently constrained the rotational
period to be $12.7047\pm0.0011$~hr at aphelion.  Using this new
period, the phase offsets between individual IRAC and MIPS
observations are minimal and range from 0.005 to 0.022 rotational
periods, with a median offset of 0.014.  Our light curve sampling of
67P differs from the approach of \citet{lamy08} who
observed the comet at 24~\micron{} with \spitzer{} 16 times---each
observation was separated by 0.5, 1.0, or 1.5~hr, spanning a total of
$\approx12.5$~hr.

\subsection{Spectra}
\begin{figure}
\includegraphics[width=\columnwidth]{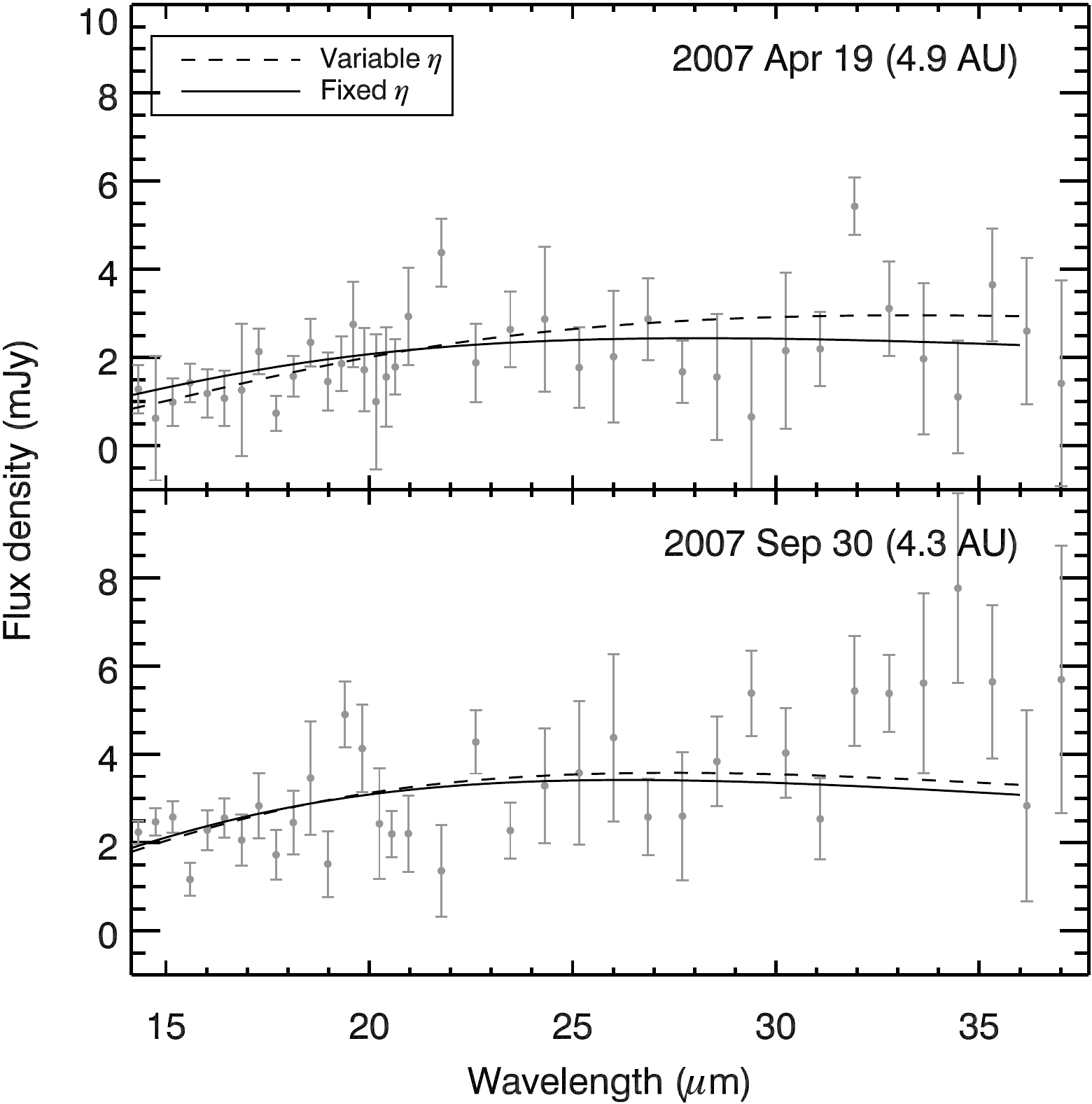}
\caption{\spitzer{} IRS 14--38~\micron{} spectra (\textit{circles}) of
  comet 67P taken at $r_h=4.9$ and 4.3~AU.  The spectral resolution
  has been reduced with a 5-point statistically weighted average.  Two
  best-fit thermal models are shown for each spectrum, one with a
  variable IR-beaming parameter ($\eta$), and one with the IR-beaming
  parameter fixed at $0.68\pm0.06$.  At 4.9~AU: $R=3.0\pm0.5$,
  $\eta=1.3\pm0.4$, $\chi^2_\nu=0.62$ (\textit{dashed-line}), and
  $R=2.11\pm0.17$, $\eta=0.68\pm0.06$, $\chi^2_\nu=0.78$
  (\textit{solid-line}).  At 4.3~AU, $R=1.99\pm0.19$,
  $\eta=0.80\pm0.14$, $\chi^2_\nu=0.62$ (\textit{dashed-line}), and
  $R=1.82\pm0.14$, $\eta=0.68\pm0.06$, $\chi^2_\nu=0.78$
  (\textit{solid-line}). \label{fig:spectra}}
\end{figure}

We obtained two spectra of comet 67P with the IRS long wavelength
(14--35~\micron), low-resolution ($\lambda/\Delta\lambda\approx100$)
module.  One spectrum was taken while 67P was 4.9~AU from the Sun, and
the other at 4.3~AU from the Sun.  The parameters of the IRS
observations are presented in Table~\ref{table:data}.  The default IRS
Stare observing template was used, which placed the comet at two
positions separated by 55\arcsec{} in the 11\arcsec{} wide by
168\arcsec{} long slit.  \spitzer{} positioned the slit at the comet's
predicted ephemeris coordinates, and followed the comet with its
computed non-sidereal rates.  The observations were repeated 9 times
for a total of 18 spectra.  We differenced the two slit positions to
remove the background, then replaced bad pixels with the average value
of the nearest neighbors or ignored them altogether.  We extracted
spectra from the images using the \spitzer{} IRS Custom Extraction
(SPICE) software package\footnote{SPICE is available at
  \url{http://ssc.spitzer.caltech.edu/}.}.  The profile of the comet
is equivalent to a point source, therefore we utilized the default
point source aperture provided by SPICE, which is $7.2\times\lambda /
27.0$~\micron~pixels wide (pixel scale 5.1\arcsec{}~pixel\inv{}).  The
individual spectra are averaged together to produce two final spectra,
one at each of 4.9~AU and 4.3~AU.  The spectra, smoothed with a
5-point statistically weighted average, are presented in
Fig.~\ref{fig:spectra}.

\noindent{\textbf{[Figure \ref{fig:spectra}]}}

\section{RESULTS AND DISCUSSION}

\subsection{The Dust Environment}\label{sec:trail}
\subsubsection{An Empirical Model}

\begin{figure}
\includegraphics[width=\columnwidth]{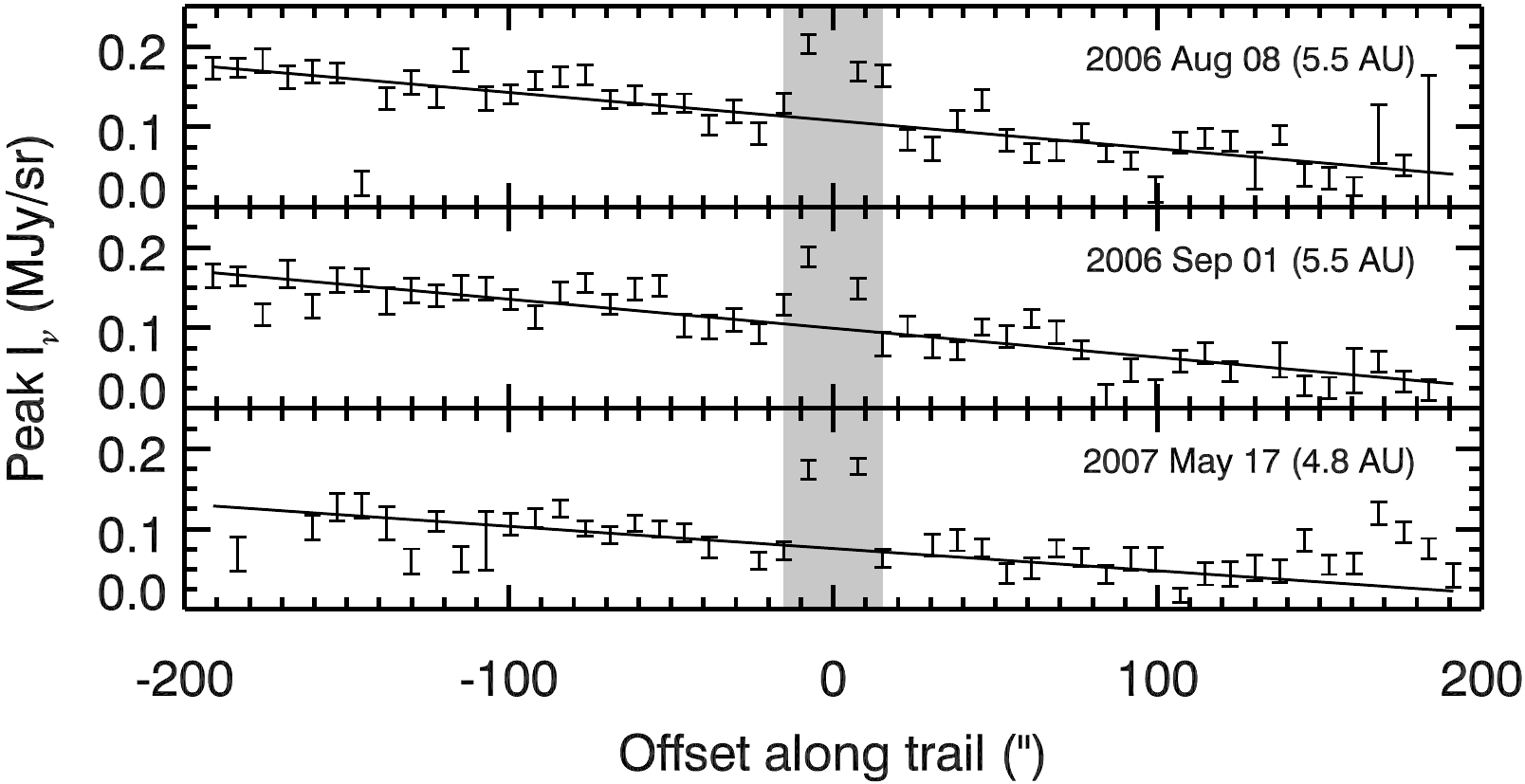}
\caption{Peak trail surface brightness and best-fit linear trends
  derived from our empirical trail model fits
  (Table~\ref{table:trail}) for three MIPS 24~\micron{} epochs.
  Positive x-axis values lie along the projected velocity vector
  (i.e., these positions are ahead of the comet nucleus).  The data
  points in the grayed area were excluded from the profile fits to
  prevent the nucleus from affecting the
  results.  \label{fig:trailpeak}}
\end{figure}

Understanding the near-nucleus dust environment of 67P when \rosetta{}
will rendezvous with the comet is important for mission planning.  Our
24~\micron{} images contain emission from large dust grains.  With
photometry of this dust, we can estimate the number density of large
grains near the nucleus, and assess their impact hazard to spacecraft.
Furthermore, if we can adequately describe the morphology of the dust
thermal emission it can be removed from the images prior to the
measurement of the comet nucleus.  To remove the dust, we fit surface
brightness profiles to linear cuts ($3\times51$~pixels) measured
perpendicular to the projected velocity vector of the comet, then
subtract our best-fit model from the images.  The assumed profile is
the sum of a Gaussian function (the dust) and a linear term (the image
background).  The results of our empirical model fits to the $r_h=4.8$
and 5.5~AU images are presented in Figure~\ref{fig:trailpeak} and
Table~\ref{table:trail}.  The photometry has been color-corrected
($\times1.057$) to the MIPS 24~\micron{} central wavelength of
23.7~\micron{} using a blackbody spectral template with
$T_d=300~r_h^{-0.5}$~K, where $T_d$ is the typical dust effective
temperature of trail grains \citep[W.~T. Reach et al., 2009, in
  preparation;][]{sykes92}.  Two trends are evident in the fits to the
dust surface brightness: 1) the peak surface brightness linearly
varies with distance from the nucleus, and 2) within the uncertainties
($\pm$ a few degrees in position angle) the dust emission is aligned
with the orbit, as would be expected for a dust trail.  To generate
model images of the dust, we approximate the morphology with the
best-fit parameters described in Table~\ref{table:trail}.  The model
dust images are subtracted from the 5.5 and 4.8~AU MIPS images.

\noindent{\textbf{[Table \ref{table:trail}, Figure \ref{fig:trailpeak}]}}

\subsubsection{A Dynamical Model}

Previous images of 67P at aphelion show a neck-line tail structure.  A
neck-line structure is a thin, linear feature caused by the projected
alignment of dust grains that were emitted at a true anomaly
180\degr{} prior to the observation in question \citep{southworth63,
  kimura75}.  The enhancement is strongest when the observer is
located near the orbital plane of the comet, which is frequently the
case for ecliptic comets such as 67P.  Small dust grains cannot remain
near the nucleus for many years; therefore, if a neck-line is observed
in our \spitzer{} observations, it must be comprised of large dust
grains that weakly interact with solar radiation pressure.
Observations of neck-lines are beneficial because they allow us to
measure the size distribution of dust grains emitted by the comet.
This becomes apparent when we consider that larger dust grains are
ejected from the nucleus with lower ejection velocities and are
accelerated less effectively by solar radiation pressure than smaller
grains.  Therefore, the largest grains remain the closest to the
nucleus, and the grain size decreases with increasing distance along
the neck-line.  In principle, the variation of the surface brightness
distribution of a neck-line with distance from the nucleus can be
transformed into a grain size distribution.  In the \spitzer{} MIPS
images, the mere existence of a neck-line can help us constrain the
largest grains emitted by 67P.  The age of a neck-line observed on
2006 August 16 would be 3.9~years, corresponding to dust emitted at a
post-perihelion distance of $r_h=1.26$~AU.  Similarly, if a neck-line
were observed on 2007 May 18, its age would be 4.6~years, i.e.,
released from the nucleus at $r_h=1.30$~AU, post-perihelion.

In order to determine if a neck-line tail structure exists in our
images, we compute simulated images of the comet using the
dynamical dust model of \citet{kelley08}.  The primary forces acting
on a dust grain are the force from solar radiation ($F_r$), the force
of solar gravity ($F_g$), and gravitational perturbations from the
planets.  Dynamical models of comets typically describe dust grains
with the parameter $\beta=F_r/F_g$.  Both forces vary with $r_h^{-2}$,
therefore $\beta$ is independent of heliocentric distance.  The
$\beta$ parameter does, however, depend on grain size:
\begin{equation}
  \beta = \frac{F_r}{F_g} = \frac{0.57 Q_{pr}}{\rho a},
\end{equation}
where $Q_{pr}$ is the radiation pressure efficiency ($\approx1$ for
large grains), $\rho$ is the grain density in units of \gcm, and $a$
is grain radius in units of \micron{} \citep{burns79}.  When
converting from $\beta$-values to grain radii, we follow
\citet{kelley08} and adopt a grain density of 1~\gcm{} (see
\citet{kelley08} for discussion related to this choice).  To describe
the dust ejection dynamics and history of comet 67P, we start with the
best parameters of \citet{kelley08}.  In summary, \citet{kelley08} fit
\spitzer{} MIPS images of the comet taken on 2004 February 23 (4.5~AU,
pre-aphelion) and found three distinct dynamical components: 1) a dust
trail with an age greater than one orbit, 2) a neck-line ejected near
the 2002 perihelion passage, and 3) a broad dust tail.  The surface
brightness distribution of the dust in the 4.5~AU, pre-aphelion, image
was best modeled by:
\begin{itemize}
\item a dust production proportional to $\cos{z_\sun}$, where
  $z_{\sun}$ is the sun-zenith angle of a surface element ejecting
  dust grains,
\item a size distribution proportional to $a^{-3.5}$,
\item an ejection velocity equal to $0.5~\beta^{0.5}~r_h^{-0.5}
  \cos{z_\sun}$~\kms, and
\item dust grains with $-5 \leq \log{\beta} \leq -1$ (6~\micron{}
  $\lesssim a \lesssim$ 60~mm).
\end{itemize}
\citet{kelley08} assumed the comet's dust production is proportional
to dust production rates inferred from optical photometry
\citet{kidger03}, $Q_d\propto r_h^{-5.8}$.  We note that the above
parameters are similar to those of \citet{ishiguro08}, derived by
comparison of dynamical models to three epochs of ground-based optical
observations.  The greatest differences between the \citet{kelley08}
and \citet{ishiguro08} models is the choice of the largest grain sizes
sizes (60~mm versus 5~mm), and dust production rates ($Q_d\propto
r_h^{-5.8}$ versus $r_h^{-3}$).

\begin{figure}
\begin{center}
  \includegraphics[width=0.9\columnwidth]{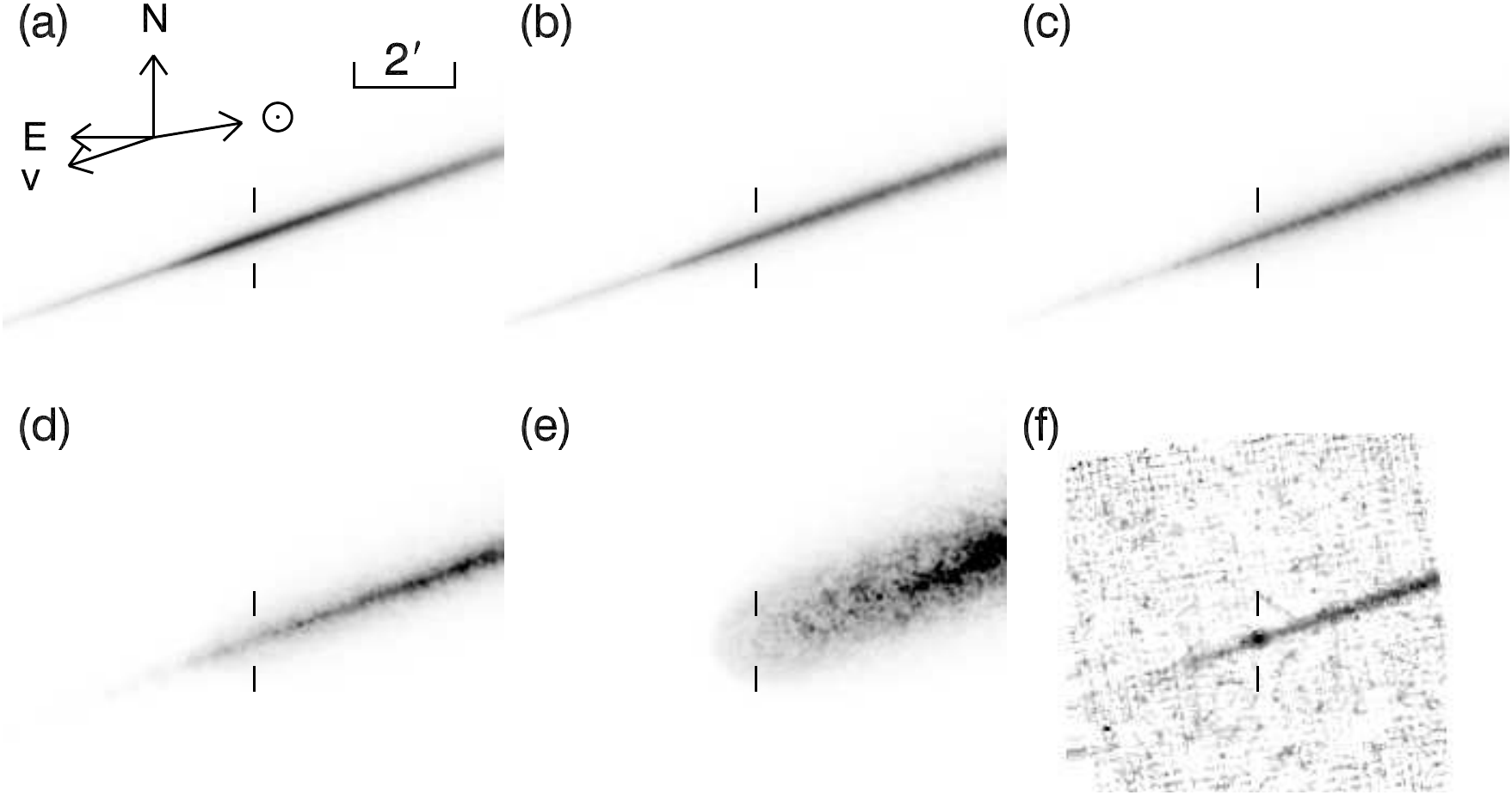}
  \includegraphics[width=0.9\columnwidth]{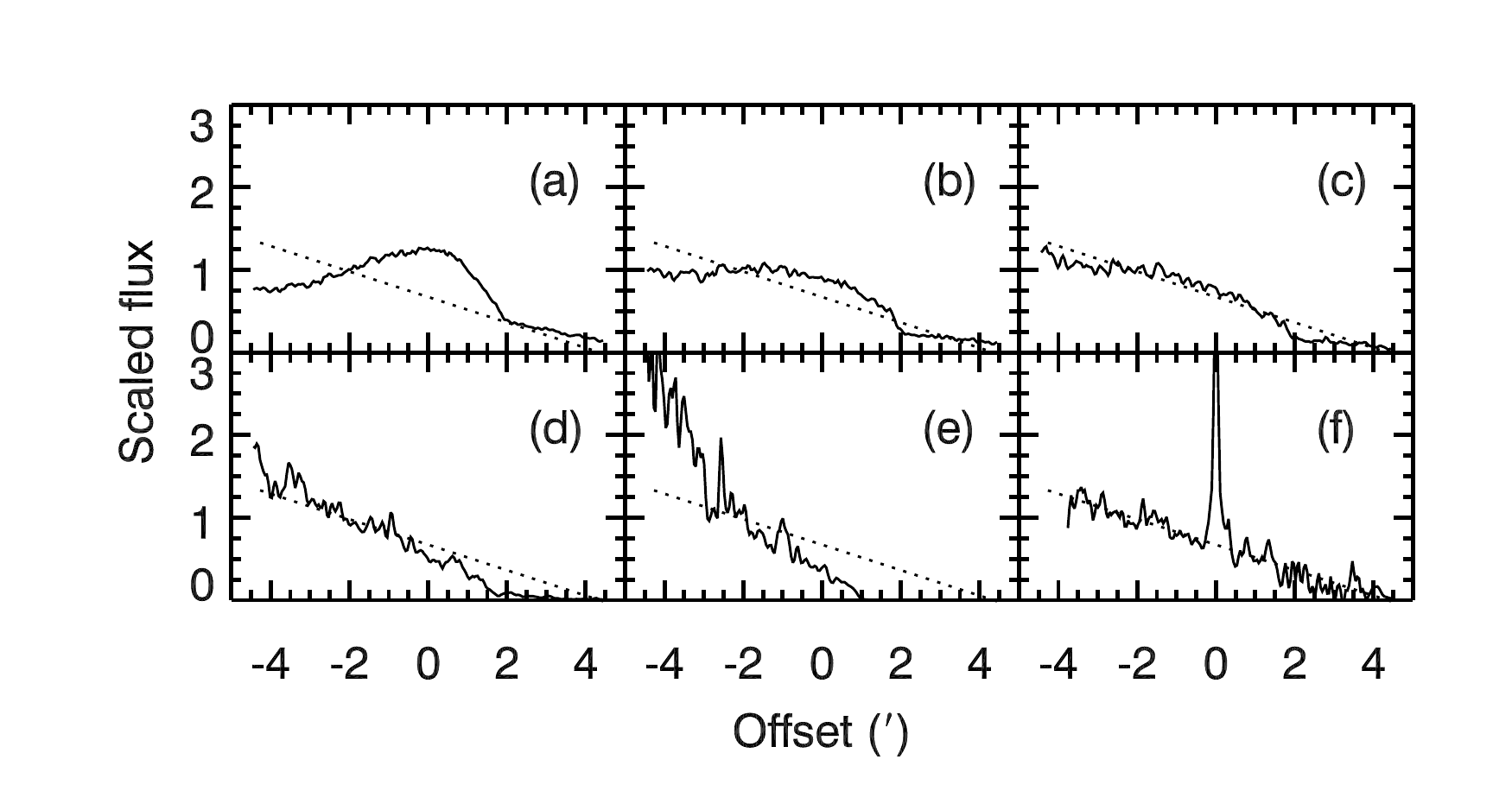}
\end{center}
\caption{Top: Simulated 24~\micron{} images (a-e) and the 2006 Aug 16
  MIPS image (f).  Each simulated image was generated with a different
  large grain cutoff: (a) $\beta>1\times10^{-5}$, (b)
  $\beta>5\times10^{-5}$, (c) $\beta>1\times10^{-4}$, (d)
  $\beta>2\times10^{-4}$, (e) $\beta>5\times10^{-4}$.  All images are
  scaled to a value of 1.0 at a distance of 2\arcmin{} behind the
  nucleus.  North is up, east is to the left, the projected Sun angle
  ($\sun$) and comet velocity vector (v) are indicated with
  \textit{arrows}, and the image scale is provided in (a).  Bottom:
  Surface brightness profiles generated from each of the simulated
  images (a-e) and the observation (f).  We fit a linear profile to
  the observed dust in (f), and plot it as a \textit{dotted-line} in
  each subplot.  \textit{Vertical lines} mark the location of the
  nucleus.  \label{fig:aug-sims}}
\end{figure}

\begin{figure}
\begin{center}
  \includegraphics[width=0.9\columnwidth]{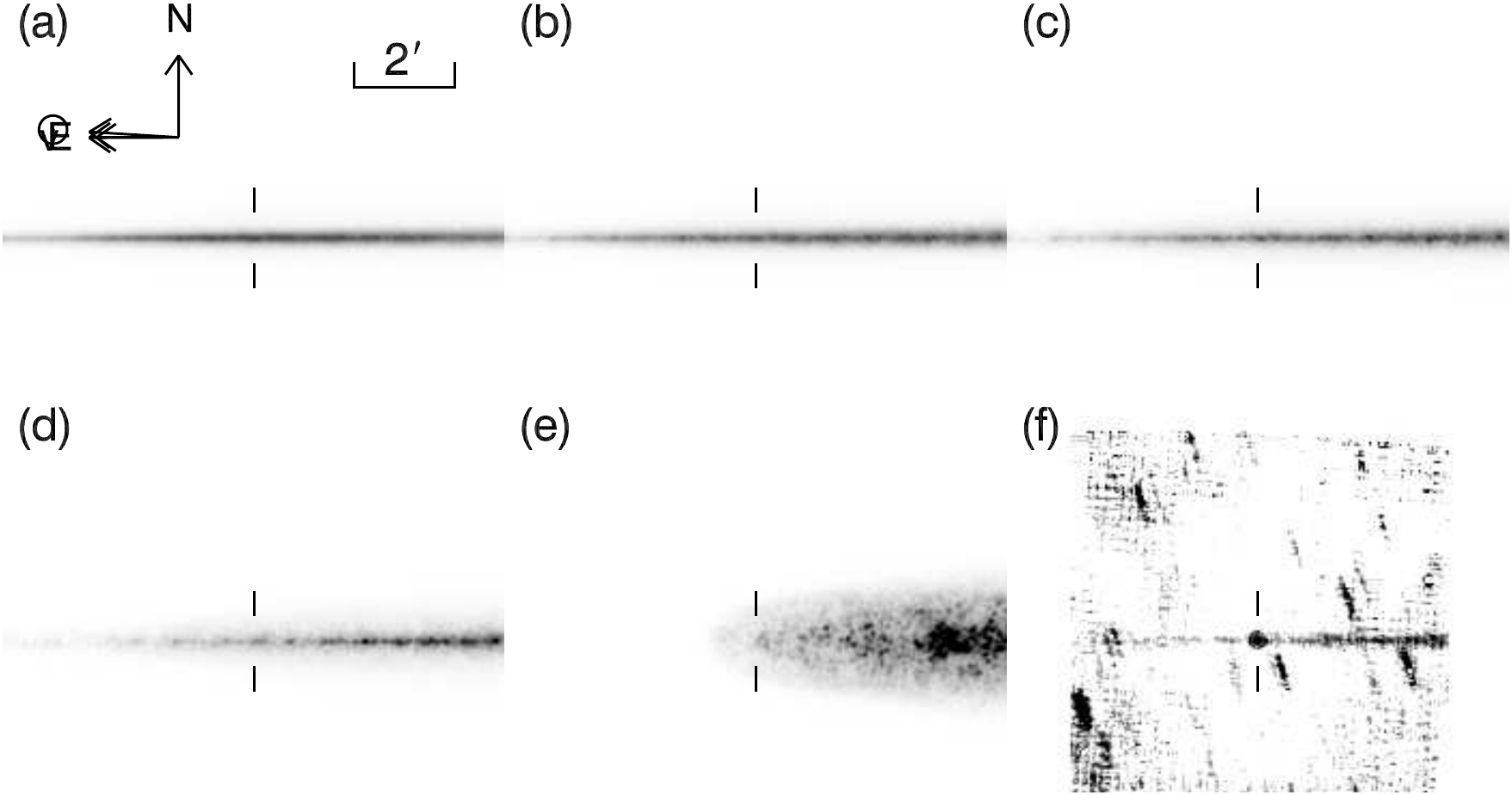}
  \includegraphics[width=0.9\columnwidth]{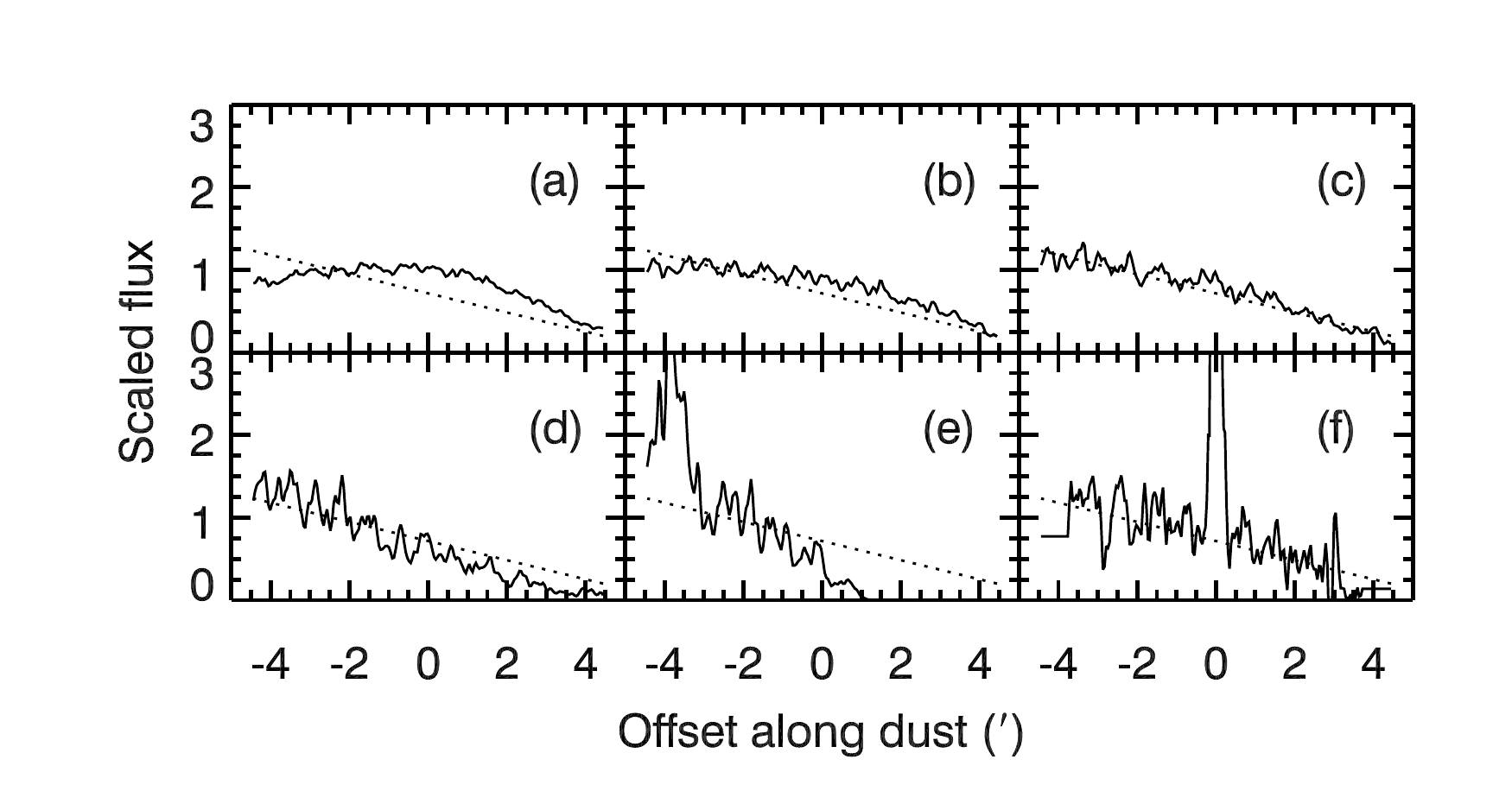}
\end{center}
\caption{Same as Fig.~\ref{fig:aug-sims}, except the 2007 May 18 epoch
  simulations and MIPS image are shown. \label{fig:may-sims}}
\end{figure}

We generated two simulations of 67P using the dynamical model and best
parameters of \citet{kelley08}, with the added stipulation that no
grains are ejected outside of $r_h=4.5$~AU.  This distance corresponds
to the comet-Sun distance during the February 2004 \spitzer{}
observations, where little to no coma was observed \citep{kelley08,
  lamy08}.  The simulations reproduce the observing conditions of our
5.5~AU and 4.8~AU images.  Each simulation tracked $\approx
2\times10^6$ grains and recorded their final positions, which are
projected onto the celestial sphere for an observer located at
\spitzer.  The oldest grains tracked were ejected near aphelion in
March 1986, and during the simulation, the comet passed perihelion
three times.  The resulting images are convolved with a model
24~\micron{} point spread function generated with version 2.0 of the
Tiny Tim/\spitzer{} software\footnote{Tiny Tim/\spitzer{} is available
  from the \spitzer{} Science Center:
  http://ssc.spitzer.caltech.edu/}.  In Fig.~\ref{fig:aug-sims} and
\ref{fig:may-sims}, we present the simulated 24~\micron{} images and
surface brightness profiles, generated with a variety of $\beta$-value
cutoffs: (a) $\beta>1\times10^{-5}$, (b) $\beta>5\times10^{-5}$, (c)
$\beta>1\times10^{-4}$, (d) $\beta>2\times10^{-4}$, (e)
$\beta>5\times10^{-4}$.  Simulations (c) and (d) best reproduce the
observed profile at both epochs.  We estimate that 67P ejects grains
with $\beta$-values larger than $1\times10^{-4}$, corresponding to
grains smaller than $\approx6$~mm in radius.  This is essentially the
same size upper-limit derived by \citet{ishiguro08}.

Formally, a neck-line tail structure is present in our \spitzer{}
observations.  The neck-line and dust trail overlap in both the 5.5~AU
and 4.8~AU images, and they cannot be resolved into two components as
has been observed by \citet{kelley08}, \citet{ishiguro08}, and
\citet{agarwal07a}.  The distinction between neck-line and trail in
our observations rests merely on their dynamical parameters, since
their grain populations have approximately the same sizes ($\sim1$~mm
in radius) at these heliocentric distances.  Taking into account our
$\beta$-value lower-limit of $10^{-4}$, the neck-line is located at a
position angle $\approx1$\degr{} north (at 5.5~AU) or south (at
4.8~AU) of the trail.  In our images, we find that 15--25\% of the
observed flux within 6.5\arcsec{} of 67P's projected orbit is
attributable to the neck-line tail, released at $r_h\approx1.3$~AU.

\subsubsection{Large Dust Grain Number Density and Impact Hazard to \rosetta{}}

\begin{figure}
\includegraphics[width=\columnwidth]{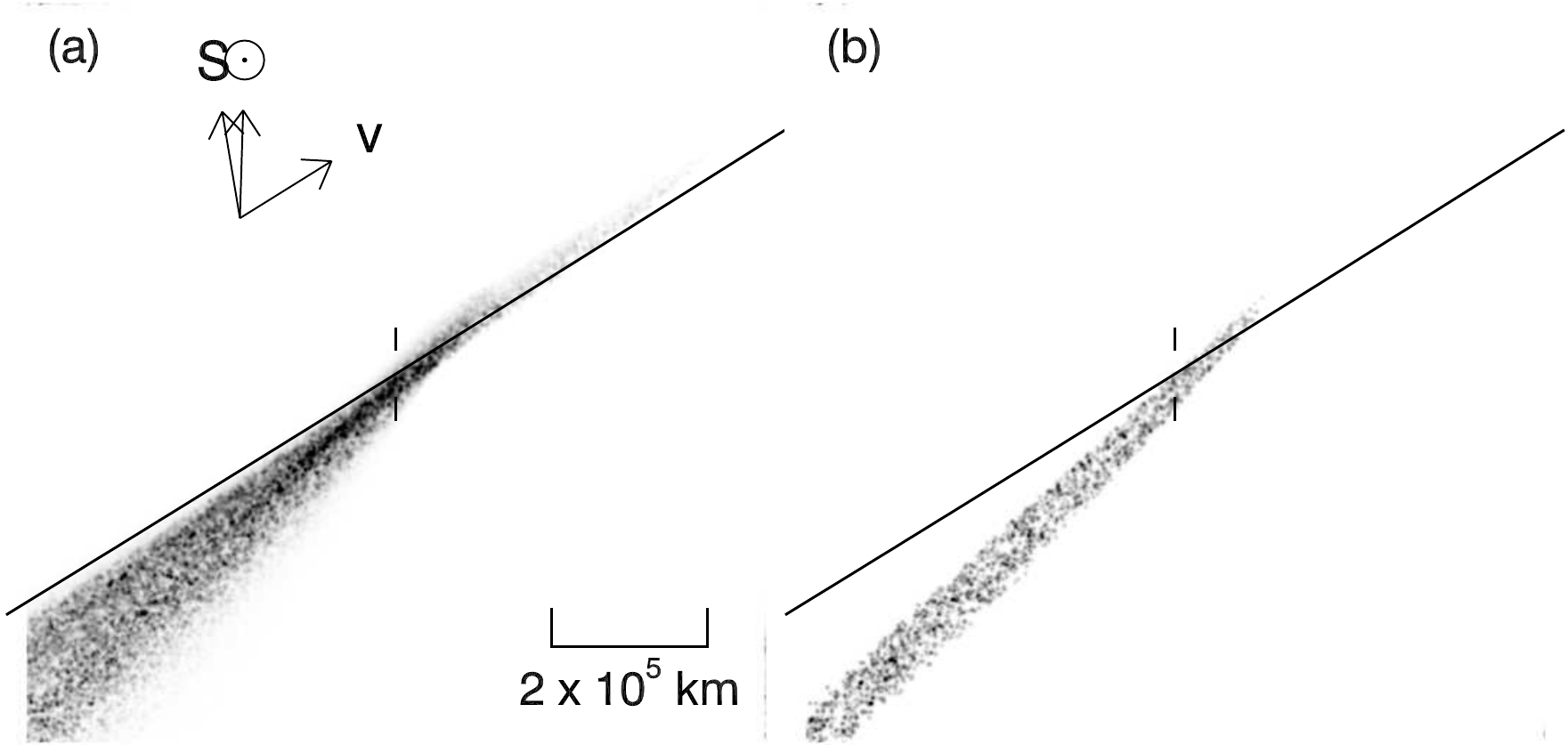}
\caption{Simulated 24~\micron{} images of 67P's 2007 May 18 dust
  environment as viewed by an observer located above the ecliptic
  plane: (a) all dust with $\beta>10^{-4}$; and (b) neck-line tail
  dust with $\beta>10^{-4}$.  A \textit{solid-line} marks the
  projected orbit of the comet, \textit{vertical lines} mark the
  location of the nucleus, the projected Sun angle ($\sun$), comet
  velocity vector (v), and \spitzer{} direction (S) are indicated with
  \textit{arrows}, and the image scale is provided in (a).  The
  surface brightness of image (b) has been scaled by a factor of 2
  relative to image (a) to improve contrast, and both images have been
  convolved with a 4600~km wide (FWHM) Gaussian kernel.  The neck-line
  tail grains and other dust ejected near perihelion have orbits that
  are distinctly different from the trail grains, which are located
  closer to the orbit of the comet. \label{fig:may-above}}
\end{figure}

\citet{kelley08} estimate the trail surface brightness near the
nucleus to be $\sim0.1$ \mjysr{} in 24~\micron{} MIPS images obtained
at 4.5~AU pre-aphelion.  Their estimate depends on a dynamical model
to separate trail emission from tail emission.  \citet{kelley08} then
derive a number density of $\sim10^{-11}$~m$^{-3}$ for millimeter
sized grains.  From our empirical model (Table~\ref{table:trail}), the
peak dust surface brightness at the nucleus is
$0.076\pm0.004$~\mjysr{} and the FWHM is $(19\pm2)\times10^3$~km at
4.8~AU, post-aphelion.  These values are improvements over the
\citet{kelley08} analysis because the dust width and surface
brightness may be measured closer to the nucleus.  Moreover, we can
estimate the extent of the dust along the \spitzer{} line-of-sight
using the above dynamical simulations.  Figure~\ref{fig:may-above}
shows the 4.8~AU simulation as viewed by an observer located above the
ecliptic plane.  The apparent width of the dust along \spitzer's
line-of-sight is $\approx60,000$~km.

\noindent{\textbf{[Figure \ref{fig:may-above}]}}

Assuming the dust is entirely comprised of millimeter sized grains
with a dust temperature of $T_d=300~r_h^{-0.5}$~K and contained within
a volume of $38,000\times60,000\times40,000$~km, we derive a grain
number density of $(1.33\pm0.06)\times10^{-12}$~m$^{-3}$.  Our assumed
volume corresponds to the observed width of the trail ($2\sigma$,
Table~\ref{table:trail}), the thickness of the simulated dust along
the line-of-sight in the $\beta>10^{-4}$ model, and the MIPS spatial
resolution.  The number density is about a factor of 10 smaller than
the estimate of \citet{kelley08}.  We caution that the uncertainty is
only based on the observed flux, and we are assuming that our chosen
simulation parameters accurately represent the behavior of the comet.

Inspection of Fig.~\ref{fig:may-above} reveals that a spacecraft, such
as \rosetta, approaching from the interior of the comet's orbit will
not encounter the bulk of the dust.  To estimate an upper-limit to
\rosetta's impact hazard based on our simulations and observations, we
assume the dust grains are uniformly distributed in space, and the
\rosetta's path length through that dust is 30,000~km (half of the
dust's thickness in comet's orbital plane).  The total impact
probability to millimeter-sized grains is $<0.003$\%~m$^{-2}$.  Taking
the spacecraft cross-section to be 4~m$^2$ and the total solar panel
cross-section to be 64~m$^2$ \citep{glassmeier07}, \rosetta{} has
$<0.3$\% chance of encountering a millimeter-sized grain during
approach to the nucleus.  During comet fly-by missions, spacecraft
typically survive collisions with a few 0.1--1~mg
($\approx500$~\micron) dust grains at high relative speeds,
$\sim1$~\kms{} \citep{ahearn05, green04, mcdonnell87, mcdonnell93}.
However, previous fly-by missions visited their respective comets
during vigorous coma activity, which enhances the density of large
grains.  Therefore, our encounter probability upper-limit is not
unusual and seems reasonable.

\subsection{Comet Activity}\label{sec:coma}

\begin{figure}
\includegraphics[width=\columnwidth]{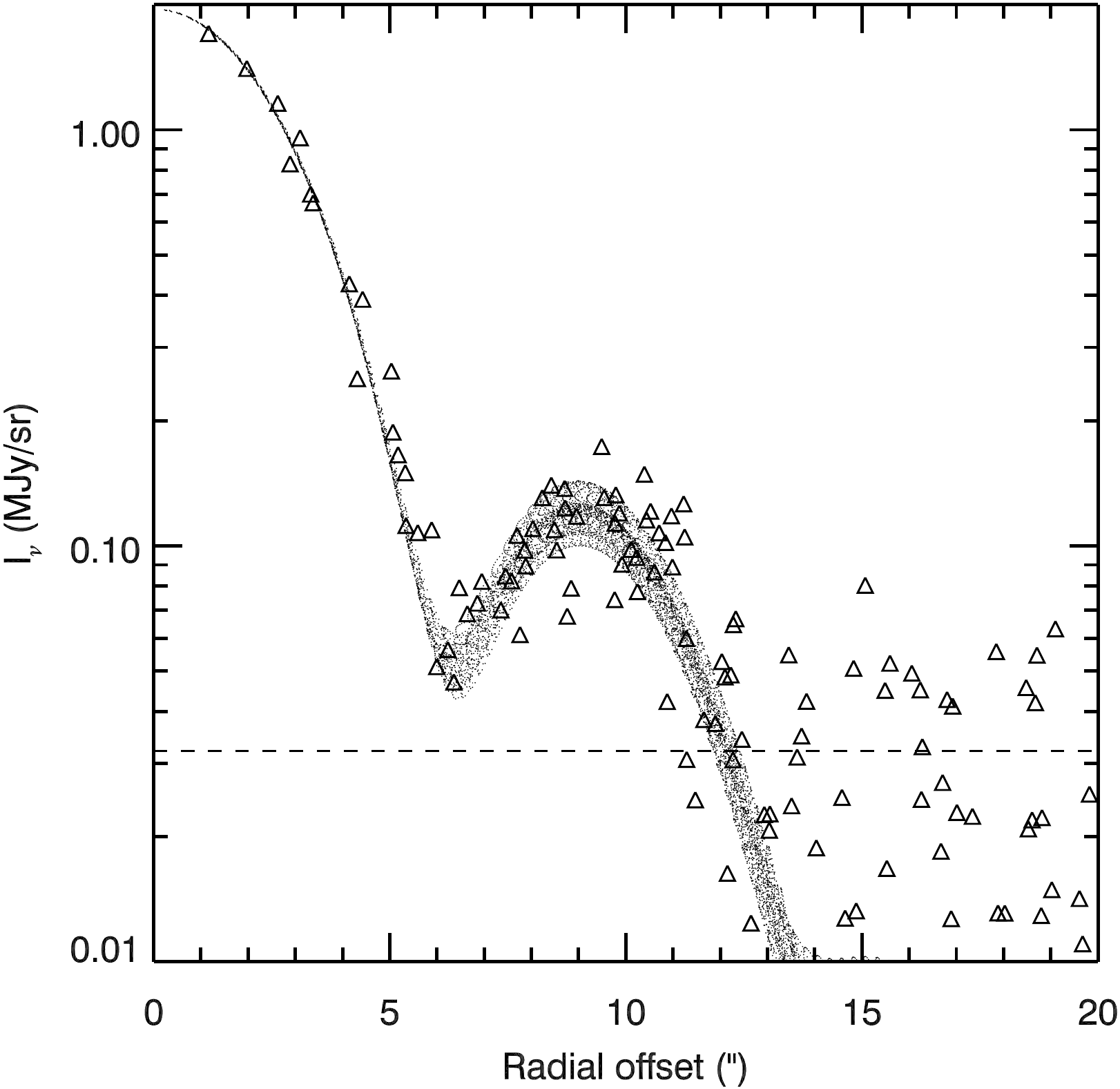}
\caption{The 24~\micron{} radial surface brightness profile
  (\textit{triangles}) of comet 67P/Churyumov-Gerasimenko at 4.8~AU
  (2007 May 18) after emission from the dust trail has been removed.
  A 200~K model point spread function is plotted with \textit{dots},
  and the $1\sigma$ background level is marked with a
  \textit{horizontal dashed-line}.  There is no evidence for a dust
  coma in this or any other image. \label{fig:psfplot}}
\end{figure}

The \spitzer{} MIPS observations are suitable for a search for faint
coma emission.  We restricted our search to the MIPS images because
they are more sensitive to emission from dust at 4--6~AU from the Sun.
First, the 11 MIPS observations at 4.8~AU are averaged together to
create a deep image, then the empirical dust model is subtracted from
the images.  Subtracting the dust model should not affect our search
for a dust coma as comae are peaked near the nucleus, yet the observed
dust varies linearly across the position of the nucleus
(Fig.~\ref{fig:trailpeak}).  The nucleus of 67P has a full-width at
half-maximum (FWHM) of 2.2~pixels, equivalent to the FWHM of point
sources in the surrounding field-of-view when the images are mosaicked
in the celestial reference frame.  A comparison of a model PSF
generated with Tiny Tim/\spitzer{} (T=200~K) to the observed radial
profile of 67P (Fig.~\ref{fig:psfplot}) strongly suggests that the
nucleus is devoid of an extensive coma at this heliocentric distance.
The azimuthally averaged coma flux at 6--8~pixels (15--20\arcsec{})
from the nucleus, i.e., outside of the first diffraction ring of a
point source, is $0.0025\pm0.0030$~\mjysr{}, yielding a $1\sigma$
upper-limit of 0.0055~\mjysr.  Similarly, we also measured a stellar
profile and a coma upper-limit of 0.0042~\mjysr in the images at
5.5~AU.

\noindent{\textbf{[Figure \ref{fig:psfplot}]}}

We can convert the surface brightness upper-limits, $I(\rho)$, to an
integrated coma brightness upper-limit if we assume a coma shape.  We
assume a steady-state coma surface brightness distribution of
\invrho{}, where $\rho$ is the angular distance from the nucleus.
This assumption is typical for photometry of distant comets
\citep[e.g.,][]{snodgrass08}.  The more steeply shaped profile of a
radiation pressure dominated coma, $I_\nu\propto\rho^{-1.5}$
\citep{jewitt87}, has a less restrictive upper-limit by a factor of 2.
The integrated coma brightness upper-limits at 24~\micron{} are
computed from $F_C \leq 2\pi\rho^2I(\rho)$, and are equal to 0.26~mJy
at $r_h=4.8$~AU, and 0.20~mJy at 5.5~AU.  These upper-limits are 9\%
of the point source fluxes presented in \S\ref{sec:nucleus}.  Removing
a 9\% coma contribution from our photometry decreases our best-fit
nucleus radii by 19\%.  In \S\ref{sec:nucleus}, we assume the coma
contribution is zero, and find that the resulting nucleus radii agree
with previous estimates.

\subsection{The Mid-infrared Light Curve and the Nucleus Size, Shape,
  and Temperature}\label{sec:nucleus}
\subsubsection{Photometry}
The MIPS and IRAC light curves allow us to constrain the surface
temperature distribution of the nucleus, which is required in order to
accurately model its size and shape.  First, we describe our
techniques for measuring the comet flux in our light curve
observations at 4.8~AU; next, we describe our techniques for the
snapshot observations at 4.4 and 5.5~AU; and finally, we describe our
nucleus thermal model and the results of the model fits to the data.

For the images taken at 4.8~AU, we subtract the empirically derived
model dust from each mosaicked MIPS observation.  Next, we subtract
the celestial background.  As the telescope tracks the comet,
background objects approach and recede from the vicinity of the
nucleus.  To remove the celestial objects, we identify image pairs
where the comet has moved at least 6 pixels.  Aligning and
differencing these image pairs removes the fixed background objects.
Finally with aperture photometry, we derive the flux of the comet
utilizing for IRAC a 5~pixel source aperture, and a 10--25~pixel
background annulus.  The aperture correction for the IRAC photometry
is $\times1.082$, and the color-correction is $\times0.869$.  For MIPS
the aperture and color-corrections are, respectively,
$\times1.180-1.349$, and $\times1.047$ for 4--6~pixel source apertures
and a 15--30~pixel background aperture.  The color-corrections were
generated and iteratively refined using a best-fit near-Earth asteroid
thermal model \citep[NEATM;][]{harris98} spectrum (see below), as
prescribed by the \citet{idh, mdh}.

At 4.4 and 5.5~AU, a slightly different approach is required.  First,
our empirical dust model is subtracted from the 5.5~AU images.  At
4.4~AU, the images were not sufficiently sensitive to detect a dust
trail, therefore no trail was subtracted.  The nucleus of 67P falls
very close to a few background stars in all images.  The comet was not
tracked sufficiently long enough to allow celestial sources to move
away from the nucleus, therefore the shift and align technique we used
to remove point source contamination in the 4.8~AU images is not
possible.  Instead, we subtract these stars with a PSF fitting program
\citep{diolaiti00}.  The program does not preserve the
critically-sampled \spitzer{} PSF adequately enough for precise
photometry (it is designed for over-sampled images), but does
sufficiently remove background stellar sources fainter than the
nucleus for our purposes.  We repeat the same photometry procedure as
executed on the 4.8~AU data (color-corrections are $\times1.046$ at
5.5~AU and $\times1.050$ at 4.4~AU).

Table~\ref{table:data} presents the color- and aperture-corrected
photometry.  The light curve photometry has also been corrected to
remove the effects of the changing observation geometry as the
distance between the Earth and 67P increased during the 1.5~days that
elapsed between the first IRAC and the last MIPS observation
($\leq1.5$\% correction).  The absolute calibration errors for MIPS
and IRAC are 3\% and 4\%, respectively, but are not included in the
reduced photometry in order to enable direct comparison of the
point-to-point variation of the data.  Absolute calibration errors are
included in the thermal model fit results discussed below.  The mean
24~\micron{} and mean 8~\micron{} fluxes at 4.8~AU are
$F_{24}=2.83\pm0.15$~mJy and $F_8=0.109\pm0.015$~mJy.

\subsubsection{Nucleus Thermal Model}\label{sec:model}

To derive 67P's nucleus size and temperature, we adopt the NEATM
\citep{harris98}, noting that this model assumes the nucleus is
spherical and in instantaneous equilibrium with insolation.  Modeling
the nuclei of comets with the NEATM has been successful in the past
\citep{fernandez06, kelley06-apj, fernandez00}, including a successful
reproduction of the global properties of the nucleus of comet
9P/Tempel substantiated by resolved temperature maps created from
\textit{Deep Impact} observations \citep{groussin07}.  Despite the
simplifications of the model, the NEATM can still provide good
estimates for the effective radius and albedo of the comet
\citep[e.g., see][]{groussin07}.

In the NEATM, the temperature distribution of the surface peaks at the
sub-solar point and decreases with the cosine of the Sun-zenith angle.
The hemisphere beyond the solar terminator has a temperature of 0~K.
The sub-solar temperature is computed via
\begin{equation}
  T_{ss} = \left[\frac{(1 - A) S}{\eta \epsilon \sigma}\right]^{1/4}\ 
    \mbox{(K)},
\label{eq:tss}
\end{equation}
where $A$ is the Bond albedo, $S=1365 r_h^{-2}$~W~m$^{-2}$ is the
solar flux at the distance of the comet ($r_h$ measured in AU), $\eta$
is the IR-beaming parameter, $\epsilon=0.9$ is the infrared
emissivity, and $\sigma$ is the Stefan-Boltzmann constant.  Commonly,
the geometric albedo, $p$, is reported in the literature, rather than
the Bond albedo \citep[$p=A/q$, where $q$ is the phase integral,
see][]{hanner81, bowell89}.  The range of geometric albedos derived
for comet nuclei is 2--6\% \citep{lamy04}, which has a minor impact on
surface temperature.  $\eta$ remains as the primary unknown quantity.
Formally, $\eta$ parameterizes how the thermal emission is
preferentially emitted toward the Sun due to surface roughness, but
because the NEATM assumes instantaneous equilibrium with insolation,
$\eta$ must also account for the thermal inertia of real world
surfaces.  In practice, $\eta$ is constrained by the shape of the
nucleus thermal emission.  We derive $\eta$ by iteratively comparing
the observed spectral energy distribution to the spectrum of a NEATM
modeled hemisphere, observed at the same phase angle.  The primary
source of uncertainty in $\eta$ is our IRAC 8~\micron{} photometry.

Since the thermal emission from comet nuclei is weakly dependent on
the surface albedo, we turn to optical measurements to constrain this
value.  We follow the methods of \citet{russell16} and
\citet{jewitt91}, to employ the relationship between radius, absolute
visual magnitude, and albedo:
\begin{equation}
  p = \left[\frac{1.496\times10^8}{R}10^{0.2(M_\sun -
    H)}\right]^2\ \mbox{(km)},
  \label{eq:pv}
\end{equation}
where $R$ is the effective radius in km, $p$ is the geometric albedo,
$H$ is the absolute optical magnitude of the nucleus, and the $M_\sun$
is the absolute magnitude of the Sun.  Using a reduced-$\chi^2$
fitting routine, we simultaneously vary $p$, $\eta$, and $R$ in
Eq.~\ref{eq:pv} and the NEATM to constrain these parameters for comet
67P.

The non-simultaneous IRAC and MIPS light curves must be phased
together using a measurement of the nucleus rotation period.  The
cadence and time span of our \spitzer{} observations do not permit
unique determination of the rotation period of the nucleus.  However,
other investigators have more appropriate observations.
\citet{lamy06} observed 67P with the \hst{} in March 2003, and
constrained the period to $P=12.41\pm0.41$~hr.  \citet{lowry06}
observed the comet over three nights with the NTT in May 2005, and
measured $P=12.72\pm0.05$~hr.  \citet{tubiana08} have phased together
VLT light curves measured in May 2006 and August 2006 (near aphelion),
and determined the period to be $12.7047\pm0.0011$~hr, commensurate
with the previously determined periods, although with greater
accuracy.  Because of the high accuracy, and that it was obtained
closest in time to our observations, we adopt the aphelion period of
12.7047~hr for our analysis of the \spitzer{} data.

The variation of the nucleus' color-temperature with rotational phase
can be assessed by combining the IRAC and MIPS light curves at 4.8~AU,
and fitting a Planck function to each pair of data points.  The
derived color-temperatures are within $1\sigma$ of the mean value,
$187\pm4$~K, at all longitudes.  Thus, we do not detect any variation
in temperature with rotation period.

As discussed in \S\ref{sec:lightcurve}, the aphelion light curves from
\citet{tubiana08} agree in amplitude with our \spitzer{} observations.
Using our mean fluxes and their mean $R$-band absolute magnitude
($H_R=15.35\pm0.04$, $M_{\sun,R}=-27.14$), we derive
$R=2.04\pm0.11$~km, $\eta=0.68\pm0.06$, and $p_R=0.054\pm0.006$, via
$\chi^2$ minimization.  We note that our fit has no degrees of freedom
because we are fitting three data points with a three parameter model.
The mean effective radius agrees with other recent measurements, which
range from 1.9 to 2.1~km once differences in albedo, absolute solar
magnitude, and adopted phase angle behaviors are taken into account
\citep{lamy07, tubiana08, tubiana08phd}.  The axial ratio, derived
from the MIPS light curve minimum (2.382~mJy) and maximum (3.37~mJy),
is $a/b\geq1.41\pm0.07$, which also agrees with the analyses of
\citet{lamy06} and \citet{tubiana08}.  The low $\eta$-value strongly
suggests the surface has a low thermal inertia, similar to the
surfaces of other comets \citep{groussin07}.  The best-fit radii from
all our MIPS observations, derived assuming $\eta=0.68\pm0.06$, are
presented in Table~\ref{table:data}.

\subsection{Comparison to Other Light Curves}\label{sec:lightcurve}

\begin{figure}
\includegraphics[width=\columnwidth]{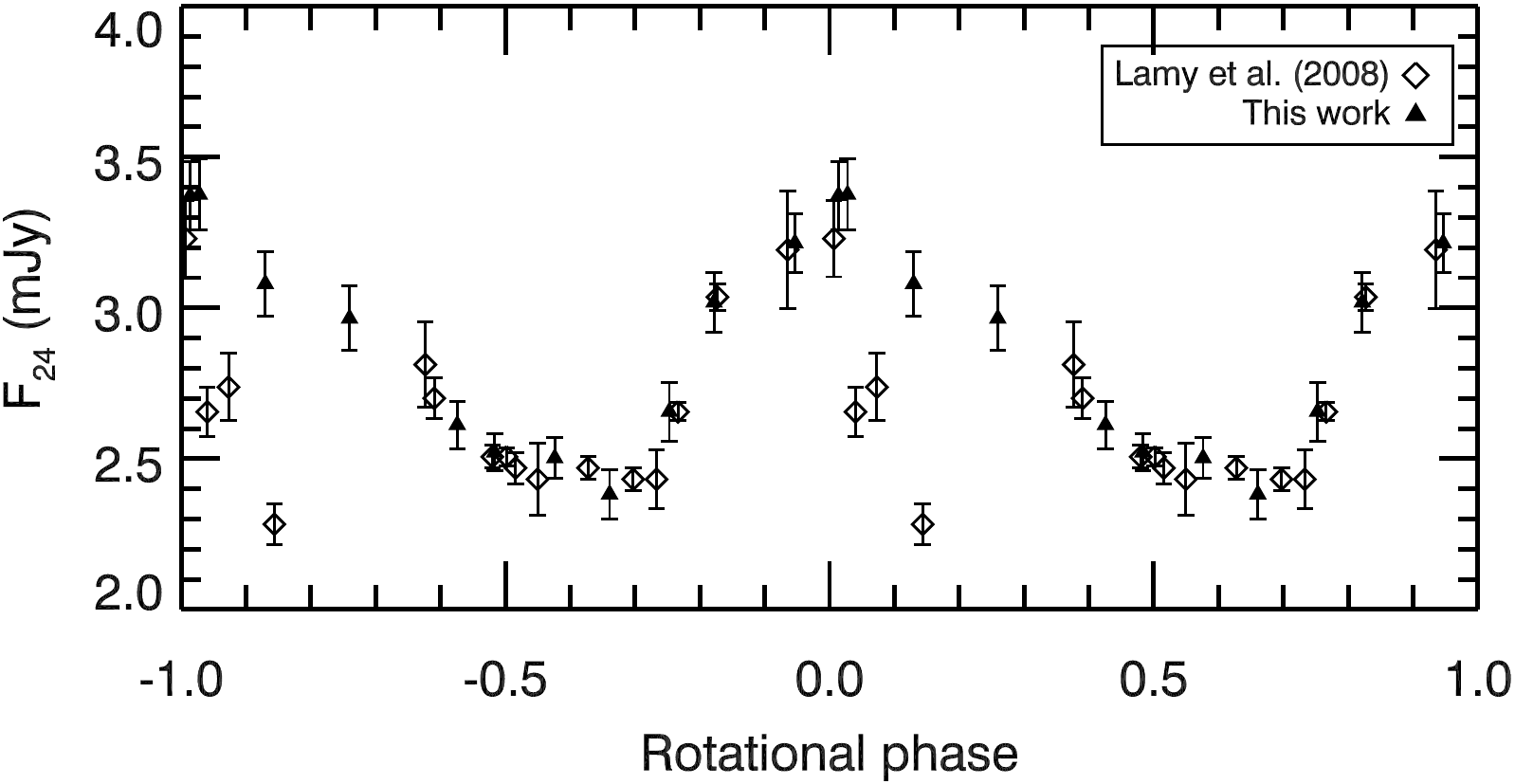}
\caption{\spitzer{} MIPS 24~\micron{} (\textit{triangles}) light curve
  (this work), and the pre-aphelion ($r_h=4.5~AU$) MIPS 24~\micron{}
  light curve from \citet{lamy08} (\textit{diamonds}).  We have phased
  the light curves with a 12.7047~hr rotational period.  The
  pre-aphelion light curve has been shifted in phase space and scaled
  by 0.78 to qualitatively match our post-aphelion
  data. \label{fig:mipslightcurve}}
\end{figure}

\citet{lamy08} created a 24~\micron{} light curve derived from
observations of 67P obtained when the comet was at 4.5~AU from the
Sun, pre-aphelion.  This latter light curve is compared to our data in
Fig.~\ref{fig:mipslightcurve}.  We scaled the \citet{lamy08} data by:
\begin{equation}
  \frac{{\rm NEATM}(r_{h,1}, \Delta_{s,1}, \alpha_{s,1}, \eta, p_R, R)}{%
    {\rm NEATM}(r_{h,2}, \Delta_{s,2}, \alpha_{s,2}, \eta, p_R, R)} = 0.746 ,
\end{equation}
where $\Delta_s$ is the comet-\spitzer{} distance, $\alpha_s$ is the
Sun-comet-\spitzer{} angle, subscript 1 designates the observing
conditions for our MIPS light curve (Table~\ref{table:data}),
subscript 2 designates the observing conditions for the \citet{lamy08}
light curve ($r_h=4.48$~AU, $\Delta_s=4.04$~AU, $\alpha_s=12.1$\degr),
$\eta=0.68$, and $p_R=0.054$.  The scale factor is independent of $R$.
The absolute phasing of the two light curves is unknown---over 2000
rotational periods have elapsed between the two observations.  We have
shifted the \citet{lamy08} light curve to minimize the point-to-point
variation between the two data sets.  Only near 0.1 phases do the two
light curves significantly disagree, but since both light curves are
sparsely sampled at these phases the discrepancies cannot be resolved.
\citet{lamy08} combine their mid-IR light curve with a shape model of
the nucleus, derived from visible data, and they estimate the $R$-band
albedo lies in the range 0.039--0.043.  This value agrees with our
best-fit value at the $2\sigma$ level, and we attribute the
discrepancy to a difference in the adopted values of the comet's
absolute optical magnitude.  Indeed, if we use the absolute magnitude
$H_R=15.46$ derived by \citet{lamy08}, our derivation of the comet's
albedo decreases to $p_R=0.049\pm0.006$, which coincides within
$1\sigma$ of the \citet{lamy08} estimates.  We prefer to adopt the
optical magnitude from \citet{tubiana08} since it was obtained when
the comet was near aphelion, similar to our \spitzer{} observations.

\begin{figure}
\includegraphics[width=\columnwidth]{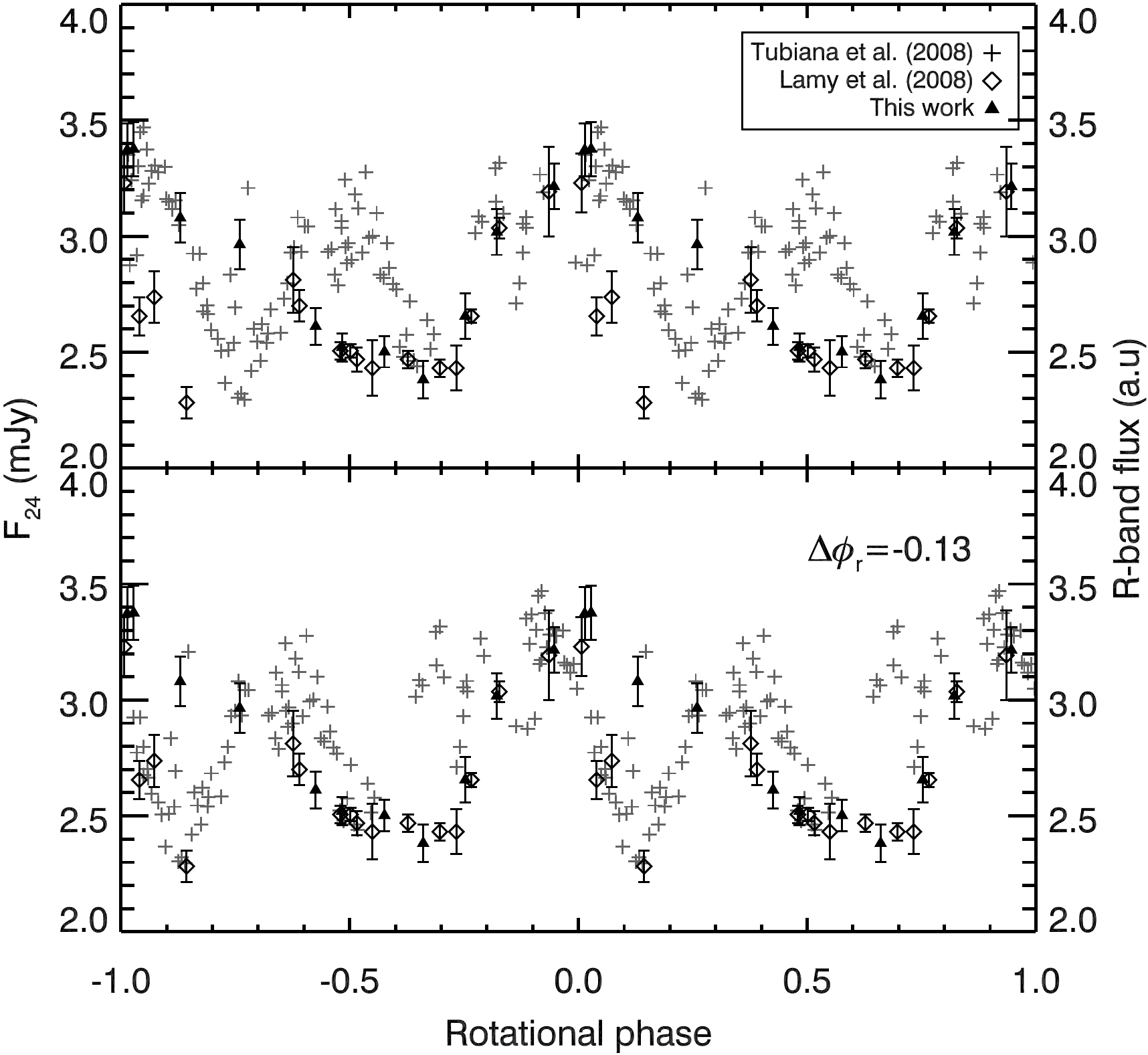}
\caption{Same as Fig.~\ref{fig:mipslightcurve}, but now the May and
  August 2006 $R$-band light curves (\textit{crosses}) from
  \citet{tubiana08} have been added.  The $R$-band light curve has
  been shifted to qualitatively match the \spitzer{} data.  Top: The
  $R$-band data has been shifted to match the mid-IR light curve
  maximum.  Bottom: The $R$-band data has been shifted by
  $\Delta\phi_r=-0.13$ phases so that the light curves better agree at
  0.3--1.0 phases. \label{fig:vislightcurve}}
\end{figure}

\noindent{\textbf{[Figures \ref{fig:mipslightcurve},
      \ref{fig:vislightcurve}]}}

Observations of 67P's visible light curve \citep{tubiana08,
  tubiana08phd} indicate that the overall variability profile is the
same throughout the aphelion arc of the cometary orbit.  Hence, we
take the well-sampled $R$-band light curve of 67P from the May and
August 2006 Very Large Telescope (VLT) data as a reference for
comparison with the \spitzer{} measurements
(Fig.~\ref{fig:vislightcurve}).  Visual inspection of the light curves
reveals that the maximum-to-minimum flux ratios (1.41 and 1.45) are
similar.  This result suggests that the variability at visible through
mid-infrared wavelengths is dominated by the effective radius of the
nucleus as it rotates, rather than variations in the albedo.

Since it is not possible to do an exact phasing of the infrared and
visible light curves, we applied arbitrary shifts to the visible light
curve in order to force the maxima into agreement
(Fig.~\ref{fig:vislightcurve}, top).  The shifted light curves have
significant differences at 0.1--0.6 phases, and are systematically
different at 0.4 to 0.6 phases.  We also attempted to minimize the
point-to-point variations between the mid-IR and $R$-band light curves
(Fig.~\ref{fig:vislightcurve}, bottom).  The overall agreement has
improved in this case, but significant differences still exist.  The
disagreements may indicate deviations in scattering and/or thermal
behaviors of the surface, e.g., variations in albedo, surface
roughness, or thermal conductivity.  Or, there may be systematic
errors in the photometry, which could explain a few rogue data points.
For example, moving sources in the background are not accounted for
with our background removal techniques, but may be faint and not
apparent in the images.  Such background sources have the greatest
effect on the faintest parts of the light curve.  It is interesting to
note that \citet{lamy08} also had some difficulty finding a perfect
agreement between their mid-IR light curve and a shape model derived
using a light curve inversion technique on optical photometry
\citep{lamy07}.  A further analysis of these discrepancies would
require detailed shape, thermophysical, and scattered light models,
better constrained phase shifts, and/or simultaneously obtained mid-IR
and visible light curves.  If we consider the overall agreements in
peak-to-peak fluxes and the approximate agreement in light curve
shapes, then we may place confidence in our nucleus albedo,
color-temperature, and effective radius computed in \S\ref{sec:model}.

\subsection{Nucleus Spectra}\label{sec:spectra}
In principle, our two spectra of the nucleus can provide an
independent measurement of the nucleus effective-temperature.  Both
spectra (Fig.~\ref{fig:spectra}) were taken during the post-aphelion
arc ($r_h=4.9$ and 4.3~AU) along with our MIPS and IRAC observations.
Using the NEATM, we fit two thermal models to each unbinned spectrum.
In one model the IR-beaming parameter is a free parameter, and in the
other model it is fixed at $\eta=0.68$ (\S\ref{sec:model}).  For both
models the albedo is fixed at $p_R=0.054$.  Our model results are
presented in Fig.~\ref{fig:spectra}, and the radii derived from the
constant $\eta$ model fits are listed in Table~\ref{table:data}.  The
absolute calibration error for the IRS instrument is approximately
10\% \citep{irsdh}, which is included in our radius uncertainties.

The IRS spectrum at $r_h=4.9$~AU does not significantly constrain the
IR-beaming parameter ($\eta=1.3\pm0.4$).  Furthermore, when the
IR-beaming parameter is allowed to vary, the best-fit radius increases
to $3.0\pm0.5$~km to compensate for the high best-fit $\eta$-value.
Such a high $\eta$-value is unphysical given the constraints from our
imaging data.  A more meaningful IR-beaming parameter and effective
radius is derived from the 4.3~AU spectrum: $R=1.99\pm0.21$~km, and
$\eta=0.80\pm0.14$.  This $\eta$-value agrees with the value of
$0.68\pm0.06$ derived from the MIPS-IRAC light curve.  Reduced
$\chi^2$ values range from 0.6--0.8 for all model fits ($\eta$-fixed
and $\eta$-variable, 4.9~AU and 4.3~AU).

\section{SUMMARY}\label{sec:summary}

We observed comet 67P/Churyumov-Gerasimenko between 5.5 and 4.3~AU
from the Sun, post-aphelion, including a full light curve of the
nucleus at 8 and 24~\micron{} at 4.8~AU.  We employ the NEATM to
derive a mean effective radius of $R=2.04\pm0.11$~km, and an average
IR-beaming parameter of $\eta=0.68\pm0.06$.  There is no evidence for
variations of the surface temperature with rotational phase.  The
amplitude of the light curve constrains the primary-to-secondary axis
ratio to $a/b\geq1.41\pm0.07$.  The mid-IR light curve is in good
agreement with the MIPS 24~\micron{} light curve presented by
\citet{lamy08}, when both datasets are phased with a 12.7047~hr
rotation period \citep{tubiana08}, and when the \citet{lamy08} light
curve is phase shifted to minimize point-to-point variations.
Together, the composite mid-IR light curve roughly agrees with two
aphelion $R$-band light curves measured by \citet{tubiana08} in May
and August 2006.  Significant discrepancies exist between the two data
sets, which may indicate surface variations of albedo, surface
roughness, or thermal conductivity.  However, the maximum-to-minimum
flux ratios agree, suggesting that the nucleus variability is
dominated by the shape of the nucleus as it rotates.  Using the
absolute magnitude $H_R=15.35\pm0.04$ \citep{tubiana08}, we estimate
the $R$-band geometric albedo to be $p_R=0.054\pm0.007$.

The \spitzer{} spectrophotometric observations of the central point
source are wholly commensurate with the thermal emission from an
inactive nucleus.  Therefore, we conclude that vigorous coma activity
did not occur before $r_h=4.3$~AU in 2007.  As the \rosetta{}
spacecraft approaches the comet during the same in-bound path in 2014,
we may expect to see images of an inactive, or weakly active nucleus.

We also observed the dust trail and large dust grain tail (i.e.,
neck-line tail structure) surrounding the nucleus in 24~\micron{}
images taken at 4.8 and 5.5~AU from the Sun.  Our other images were
not sensitive enough to detect the dust.  Starting with the dust
ejection parameters of \citet{kelley08} and \citet{ishiguro08}, and
additionally imposing a maximum grain size of 6~mm ($\beta>10^{-4}$),
simulations of the aphelion dust environment are consistent with our
MIPS 24~\micron{} observations.  Our maximum grain size agrees with
the maximum grain size of 5~mm derived by \citet{ishiguro08}.  The
grain number density near the nucleus is
$(1.33\pm0.03)\times10^{-12}$~m$^{-3}$, assuming the dust is composed
of millimeter-sized grains.  The density corresponds to a \rosetta{}
dust impact probability upper-limit of 0.3\% during the spacecraft's
approach to the nucleus in 2014.

\acknowledgments The authors acknowledge the Director of the
\spitzer{} Science Center for providing Director's Discretionary Time,
and appreciate the hard work of Vikki Meadows and the Observing
Support Staff at the SSC for scheduling our rotationally phased light
curve observations, making possible the surface color-temperature
measurements.  This research made use of Tiny Tim/\spitzer{},
developed by John Krist for the \spitzer{} Science Center.  This work
is based in part on observations made with the \sst{}, which is
operated by the Jet Propulsion Laboratory, California Institute of
Technology under a contract with NASA.  Support for this work was
provided by NASA through contract 1289123 issued by JPL/Caltech to the
University of Central Florida, and contracts 1263741, 1256406, and
1215746 issued by JPL/Caltech to the University of Minnesota.
C.E.W. also acknowledges support from the National Science Foundation
grant AST-0706980.

Facilities: \facility{Spitzer}

\renewcommand\arraystretch{0.9}
\begin{deluxetable}{llc r@{.}l r@{.}l r@{.}l r@{.}l r@{.}l r@{.}l@{\,$\pm$\,}r@{.}l r@{.}l@{\,$\pm$\,}r@{.}l}
\small
\rotate
\tablecolumns{21}
\tablewidth{0pt}

\tablecaption{Comet 67P/Churyumov-Gerasimenko observation log,
  photometry, and effective
  radii.\tablenotemark{a,b}\label{table:data}}

\tablehead{
  \colhead{AOR Key}
  & \colhead{Date}
  & \colhead{Time}
  & \multicolumn{2}{c}{Duration}
  & \multicolumn{2}{c}{$\phi_R$}
  & \multicolumn{2}{c}{$r_h$}
  & \multicolumn{2}{c}{$\Delta_s$}
  & \multicolumn{2}{c}{$\alpha_s$}
  & \multicolumn{4}{c}{$F_\nu$}
  & \multicolumn{4}{c}{$R$} \\
  & \colhead{(UT)}
  & \colhead{(hh:mm:ss)}
  & \multicolumn{2}{c}{(min)}
  & \multicolumn{2}{c}{}
  & \multicolumn{2}{c}{(AU)}
  & \multicolumn{2}{c}{(AU)}
  & \multicolumn{2}{c}{(deg)}
  & \multicolumn{4}{c}{(mJy)}
  & \multicolumn{4}{c}{(km)}
}

\startdata
\cutinhead{IRAC 7.87~\micron{}}
21631232 & 2007 May 16 & 23:55:28 & 13&6 & 0&000 & 4&852 & 4&407 & 11&22 & 0&124 & 0&021 & \blnk & \blnk \\
21630976 & 2007 May 17 & 01:21:41 & 13&6 & 0&113 & 4&852 & 4&406 & 11&22 & 0&134 & 0&014 & \blnk & \blnk \\
21630464 & 2007 May 17 & 03:21:00 & 13&6 & 0&270 & 4&852 & 4&404 & 11&21 & 0&115 & 0&014 & \blnk & \blnk \\
21630208 & 2007 May 17 & 05:14:31 & 13&6 & 0&419 & 4&852 & 4&403 & 11&21 & 0&096 & 0&013 & \blnk & \blnk \\
21629952 & 2007 May 17 & 05:56:08 & 13&6 & 0&473 & 4&851 & 4&402 & 11&21 & 0&104 & 0&014 & \blnk & \blnk \\
21629696 & 2007 May 17 & 07:14:17 & 13&6 & 0&576 & 4&851 & 4&401 & 11&20 & 0&108 & 0&013 & \blnk & \blnk \\
21629440 & 2007 May 17 & 08:07:51 & 13&6 & 0&646 & 4&851 & 4&400 & 11&20 & 0&085 & 0&010 & \blnk & \blnk \\
21629184 & 2007 May 17 & 09:25:59 & 13&6 & 0&748 & 4&851 & 4&399 & 11&20 & 0&108 & 0&014 & \blnk & \blnk \\
21631744 & 2007 May 17 & 10:19:31 & 13&6 & 0&819 & 4&851 & 4&399 & 11&19 & 0&100 & 0&014 & \blnk & \blnk \\
21630720 & 2007 May 17 & 11:44:04 & 13&6 & 0&930 & 4&851 & 4&398 & 11&19 & 0&120 & 0&014 & \blnk & \blnk \\
21631488 & 2007 May 17 & 12:52:03 & 13&6 & 1&019 & 4&850 & 4&397 & 11&19 & 0&119 & 0&014 & \blnk & \blnk \\
\cutinhead{MIPS 23.7~\micron{}}
18555904 & 2006 Aug 16 & 02:16:18 & 30&3 & \blnk & 5&514 & 4&937 &  9&30 & 1&995 & 0&049 & 2&18 & 0&18 \\
18556416 & 2006 Sep 01 & 01:56:47 & 30&3 & \blnk & 5&488 & 5&149 & 10&44 & 1&750 & 0&051 & 2&08 & 0&18 \\
21628928 & 2007 May 18 & 01:34:42 & 12&7 & 2&019 & 4&849 & 4&387 & 11&15 & 3&37  & 0&11  & 2&19 & 0&17 \\
21628672 & 2007 May 18 & 03:03:02 & 12&7 & 2&135 & 4&848 & 4&386 & 11&15 & 3&08  & 0&11  & 2&12 & 0&16 \\
21628160 & 2007 May 18 & 04:42:04 & 12&7 & 2&265 & 4&848 & 4&385 & 11&14 & 2&97  & 0&11  & 2&06 & 0&16 \\
21627904 & 2007 May 18 & 06:48:40 & 12&7 & 2&431 & 4&848 & 4&383 & 11&14 & 2&612 & 0&079 & 1&93 & 0&15 \\
21627648 & 2007 May 18 & 07:32:44 & 12&7 & 2&489 & 4&848 & 4&383 & 11&14 & 2&521 & 0&061 & 1&91 & 0&15 \\
21627392 & 2007 May 18 & 08:43:23 & 12&7 & 2&582 & 4&848 & 4&382 & 11&13 & 2&502 & 0&067 & 1&90 & 0&15 \\
21627136 & 2007 May 18 & 09:47:50 & 12&7 & 2&666 & 4&848 & 4&381 & 11&13 & 2&382 & 0&081 & 1&83 & 0&14 \\
21626880 & 2007 May 18 & 10:58:06 & 12&7 & 2&758 & 4&847 & 4&380 & 11&13 & 2&656 & 0&098 & 1&95 & 0&15 \\
21626624 & 2007 May 18 & 11:50:40 & 12&7 & 2&827 & 4&847 & 4&379 & 11&12 & 3&018 & 0&098 & 2&05 & 0&16 \\
21628416 & 2007 May 18 & 13:25:44 & 12&7 & 2&952 & 4&847 & 4&378 & 11&12 & 3&215 & 0&098 & 2&14 & 0&17 \\
21632000 & 2007 May 18 & 14:27:39 & 12&7 & 3&033 & 4&847 & 4&377 & 11&12 & 3&38  & 0&12  & 2&18 & 0&17 \\
18557440 & 2007 Sep 19 & 01:32:42 &  7&5 & \blnk & 4&376 & 3&852 & 12&32 & 5&13  & 0&41  & 2&25 & 0&19 \\
18557696 & 2007 Sep 19 & 01:41:56 &  7&5 & \blnk & 4&376 & 3&852 & 12&32 & 4&62  & 0&45  & 2&22 & 0&19 \\
18557952 & 2007 Sep 19 & 01:51:10 &  7&5 & \blnk & 4&376 & 3&852 & 12&33 & 4&70  & 0&54  & 2&16 & 0&20 \\
\cutinhead{IRS 14--38 ~\micron{}}
18555136 & 2007 Apr 19 & 05:27:36 & 90&2 & \blnk & 4&942 & 4&936 & 11&67 & \blnk & \blnk & 2&11 & 0&20 \\
18555392 & 2007 Sep 30 & 18:19:58 & 90&2 & \blnk & 4&325 & 3&974 & 13&34 & \blnk & \blnk & 1&82 & 0&17 \\
\enddata

\tablenotetext{a}{Table columns: AOR Key, \spitzer{} astronomical
  observation request identifier; Time, start of observation;
  Duration, length of observation; $\phi_R$, rotational phase at
  4.8~AU, using a 12.7047~hr rotation period; $r_h$, heliocentric
  distance; $\Delta_s$, comet-\spitzer{} distance; $\alpha_s$,
  Sun-comet-\spitzer{} angle; $F_\nu$, flux density; and $R$,
  effective radius using the NEATM with an IR-beaming parameter of
  $0.68\pm0.06$.}

\tablenotetext{b}{Instrument absolute calibration errors (3\% for
  IRAC, 4\% for MIPS, and 10\% for IRS) are not included in the
  photometric errors to facilitate point-to-point comparisons.
  Calibration errors are incorporated into the effective radius
  error.}

\end{deluxetable}
\renewcommand\arraystretch{1.0}

\begin{deluxetable}{lcccccc}
\small
\rotate
\tablecolumns{7}
\tablewidth{0pt}

\tablecaption{Empirical model parameters used to describe the dust
  trail morphology and brightness.\tablenotemark{a}\label{table:trail}}

\tablehead{
  \colhead{Observation}
  & \multicolumn{2}{c}{$A$ (\mjysr)\tablenotemark{b}}
  & \multicolumn{2}{c}{$\mu$ ($10^3$~km)}
  & \multicolumn{2}{c}{$\sigma$ ($10^3$~km)} \\ \cline{2-3}
  \cline{4-5} \cline{6-7}
  \colhead{Date}
  & \colhead{$m$} & \colhead{$b$}
  & \colhead{$m$} & \colhead{$b$}
  & \colhead{$m$} & \colhead{$b$}
}

\startdata
2006 Aug 08 (5.5~AU) &
   $(-3.5\pm0.2)\times10^{-4}$ &
   $0.108\pm0.002$ &
   0.0 &
   $-1.5\pm2.1$ &
   0.0 &
   $24\pm1$ \\

2006 Sep 01 (5.5~AU) &
   $(-3.6\pm0.2)\times10^{-4}$ &
   $0.100\pm0.002$ &
   0.0 &
   $-1.5\pm1.1$ &
   0.0 &
   $24\pm2$ \\

2007 May 18 (4.8~AU) &
   $(-2.8\pm0.3)\times10^{-4}$ &
   $0.076\pm0.002$ &
   0.0 &
   $-1.0\pm1.6$ &
   0.0 &
   $19\pm2$ \\
\enddata

\tablenotetext{a}{The model trail is a Gaussian function: $I(x,y) =
  A(y) \exp{[(x - \mu(y))^2 / \sigma(y)^2]}$, where $I$ is the surface
  brightness in units of \mjysr, $A$ is the peak surface brightness in
  units of \mjysr, $\mu$ is the trail offset from the projected orbit
  in units of km, $\sigma$ is the trail width in units of km, and $x$
  and $y$ are, respectively, offsets perpendicular and parallel to the
  projected orbit, measured in units of arcseconds.  Each parameter,
  $A$, $\mu$, and $\sigma$, is a linear function, e.g., $A=my+b$.}

\tablenotetext{b}{The uncertainties in parameter $A$ do not include
  the MIPS absolute calibration error of 4\%.}

\end{deluxetable}

\begin{figure}
\includegraphics[width=\columnwidth]{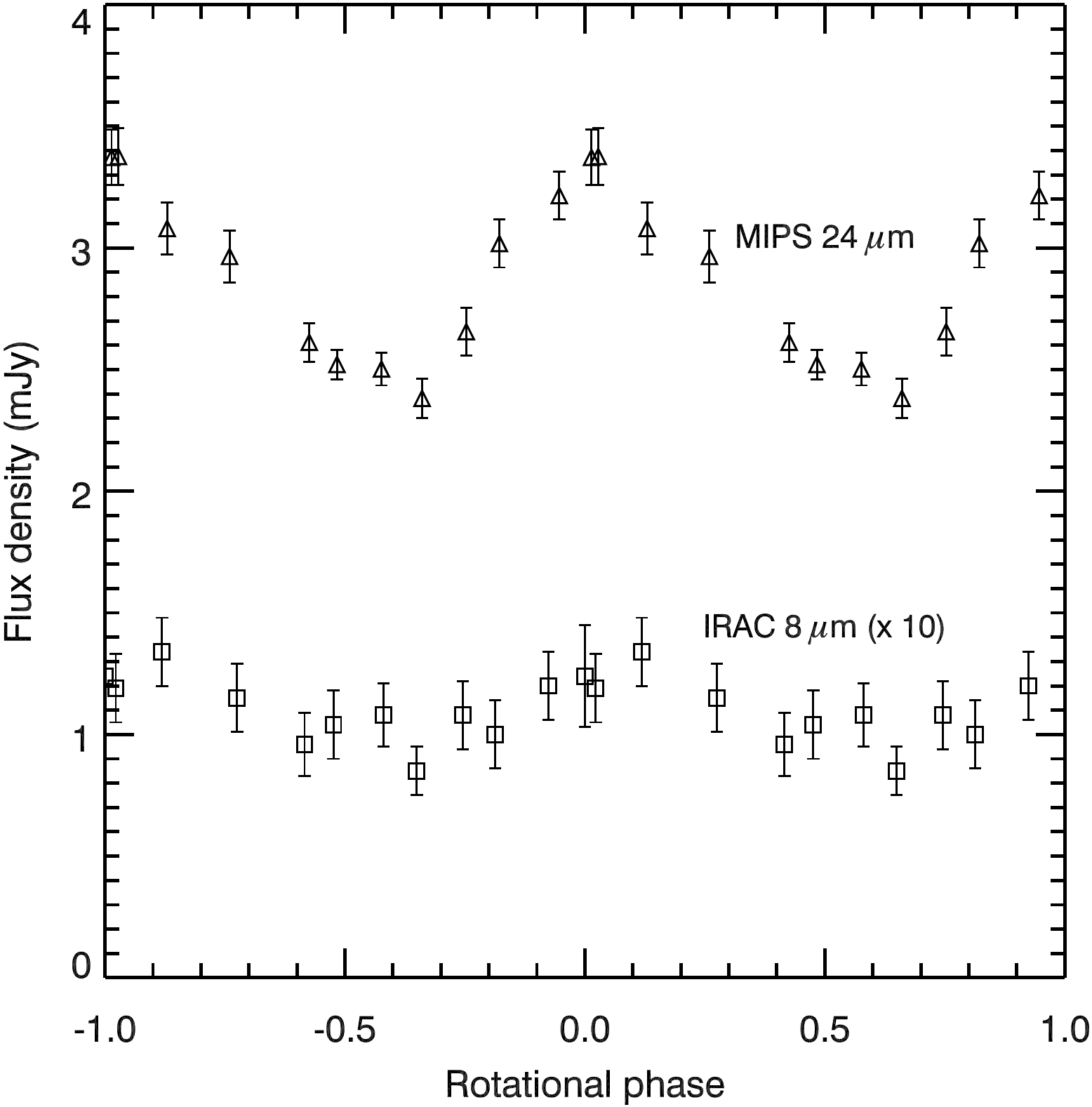}
\caption{\spitzer{} MIPS 24~\micron{} (\textit{triangles}) and IRAC
  8~\micron{} (\textit{squares}) light curves, phased with a
  rotational period of 12.7047~hr.  The IRAC light curve has been
  scaled by a factor of 10. \label{fig:mipsiraclightcurve}}
\end{figure}

\end{document}